\RequirePackage{fix-cm}
\documentclass[twocolumn]{svjour3}          
\smartqed  
\usepackage{graphicx}
\usepackage{algorithm}
\usepackage{algorithmic}
\usepackage{mathtools}
\usepackage{amssymb}
\usepackage{mathptmx}
\usepackage{xcolor}

\usepackage[bookmarks=true, bookmarksnumbered=true, bookmarksopen=true,	
		unicode=true, colorlinks=true,
		pagebackref=true,
		linkcolor=blue,citecolor=blue,filecolor=blue,urlcolor=blue
		]{hyperref}
\begin{document}

\title{Interpreting communities based on the evolution of a dynamic attributed network
}

\author{G\"{u}nce Keziban Orman \and
        Vincent Labatut \and 
        Marc Plantevit \and
        Jean-Fran\c{c}ois Boulicaut
}

\institute{G\"{u}nce Keziban Orman  \at
              Computer Engineering Department, Galatasaray University \\
              \email{korman@gsu.edu.tr}            
           \and
           Vincent Labatut \at
                         Laboratoire Informatique d'Avignon, Universit\'{e} d'Avignon\\
                         \email{vincent.labatut@univ-avignon.fr}         
                      \and
 		 Marc Plantevit \at
                         Universit\'{e} Lyon 1, CNRS, LIRIS \\
                         \email{marc.plantevit@liris.cnrs.fr}           
                      \and                      
 	Jean-Fran\c{c}ois Boulicaut \at
                        INSA-Lyon, CNRS, LIRIS \\
                         \email{jean-francois.boulicaut@insa-lyon.fr}           
}

\date{Received: date / Accepted: date}

\maketitle
\sloppy 
\begin{abstract}
Many methods have been proposed to detect communities, not only in plain, but also in attributed, directed or even dynamic complex networks. From the modeling point of view, to be of some utility, the community structure must be characterized relatively to the properties of the studied system. However, most of the existing works focus on the detection of communities, and only very few try to tackle this interpretation problem. Moreover, the existing approaches are limited either by the type of data they handle, or by the nature of the results they output. In this work, we see the interpretation of communities as a problem independent from the detection process, consisting in identifying the most characteristic features of communities. We give a formal definition of this problem and propose a method to solve it. To this aim, we first define a sequence-based representation of networks, combining temporal information, community structure, topological measures, and nodal attributes. We then describe how to identify the most emerging sequential patterns of this dataset, and use them to characterize the communities. We study the performance of our method on artificially generated dynamic attributed networks. We also empirically validate our framework on real-world systems: a DBLP network of scientific collaborations, and a LastFM network of social and musical interactions.
\keywords{Dynamic Attributed Networks\and Community Interpretation \and Topological Measures \and Emerging Sequence Mining}
\end{abstract}

\section{Introduction}
\label{sec:intro}
Complex networks have become very popular as a modeling tool during the last decade, because they help to better understand the intrinsic laws and dynamics of complex systems. A typical \textit{plain} network contains only nodes and links between them, but it is possible to enrich it with different types of data: link orientation and/or weight, temporal dimension, attributes describing the nodes or links, etc. This flexibility allowed to use complex networks to study real-world systems in many fields: sociology, physics, genetics, computer science, etc. \cite{Newman2003}.

The complex nature of the modeled systems leads to the presence of non-trivial topological properties in the corresponding networks. Among them, the community structure is one of the most studied. The notion of \textit{community} originally comes from social sciences. It traditionally refers to groups of persons sharing a common territory (neighborhood, town, city, etc.) or having common relationships (human relationship, family, etc.) \cite{Gusfield1975}. More recently, it has been used to point out at groups of persons sharing emotions, having a feeling of belonging together \cite{McMillanJanuary1986}. In network science, a community roughly corresponds to a group of nodes more densely interconnected, relatively to the rest of the network \cite{Fortunato2010}. The \textit{community structure} of a network denotes to the way its communities are interconnected. Such a structure has been observed in many real-world networks \cite{Newman2003}, and it was shown to be directly related to the way the modeled systems works \cite{Fortunato2010}. It is therefore widely studied, for many objectives: discovering functionally related objects, studying interactions between modules, inferring missing attribute values and predicting unobserved connections, etc. \cite{Yang2013}. The applications are numerous, such as: recommendation systems \cite{chen2014}, viral marketing \cite{Leskovec2007} or sentiment analysis \cite{Parau2013}.

Because of this popularity, hundreds of different algorithms were developed for community detection \cite{Fortunato2010}. Although these methods differ in terms of nature of the detected communities, type of network they handle, technique of detection, algorithmic complexity, result quality and other aspects, their output can always be basically described as a list of node groups. More specifically, in the case of mutually exclusive communities, it is a partition of the set of nodes. From an applicative point of view, the question is then to make sense of these groups relatively to the studied system. In other words, for the community structure to be useful, it is necessary to interpret the detected communities. This problem is extremely important from the end user's perspective. And yet, almost all works in the field of community detection concern the definition of detection tools, and their evaluation in terms of performance \cite{Fortunato2009}. Few researchers have addressed the problem of characterizing and interpreting the communities \cite{Tumminello2011,Labatut2012,Labatut2013,Yang2013}. The existing methods suffers from various limitations: some are subjective, others focus on nodal attributes only, or on topological properties only, or mix these data in an unclear way, as we explain in more details in Section \ref{subsec:relatedworks}. Moreover, in most of these works, the problem of community interpretation is not defined as a problem in itself.

In this work, we consider the interpretation problem as independent from the approach used for community detection. We break it down to two separate subproblems: on the one hand, representing a community in an appropriate way, and on the other hand, finding the most characteristic elements from this representation. To solve them, we propose an approach based on the original definition of the notion of community in social sciences, which underlines that nodes belonging to the same community should be relatively similar and/or share a common behavior. Assessing node similarity requires describing nodes, which can be performed both in terms of \textit{individual} information (i.e. personal characteristics) and \textit{relational} information (i.e. connection to the rest of the network). Concretely, the former corresponds to nodal attributes, whereas the latter depends on the network topology. The behavior of a node can be described in terms of evolution of its individual and relational information. To take these three aspects (individual, relational, temporal) into account, we need to work with dynamic attributed networks, i.e. time evolving networks whose nodes are described with various fields. to summarize, our aim is to detect common changes in topological features and attribute values over time periods, in dynamic attributed networks. More precisely, we aim at finding the most characteristic sequential patterns for each community. These represent the general trends for the community, and provide a support for its interpretation.

Our first contribution is to formalize community interpretation as a specific problem, distinct from community detection. In particular, it should be independent from the method used to detect communities, rely on an easily replicable systematic approach, and be as automated as possible. Our second contribution is the definition of a method taking advantage of a sequential representation of networks, in order to extract characteristic patterns allowing the interpretation of communities. Our third contribution is to evaluate our method on both artificially generated and real-world networks. To this aim, we propose an extension of an existing generative model \cite{Greene2010}, in order to produce attributed dynamic networks. The real-world data are an existing co-authorship network coming from DBLP \cite{Desmier2012} and a network of Jazz listeners we extracted from LastFM.

The rest of this article is organized as follows. In Section~\ref{subsec:relatedworks}, we review and comment in more details the existing works more or less directly related to community interpretation. Then, in Section~\ref{sec:overview} we give a general description of the approach we propose to solve this problem, and illustrate it on a small toy network. In Section~\ref{sub:Definition}, we introduce some necessary concepts related to networks analysis and sequential pattern mining, before giving a formal description of the community interpretation problem. Our method for community characterization is then described in details in Section~\ref{sec:method}. In Sections~\ref{sec:Eval} and \ref{sec:Valid}, we present our results obtained on artificial and real-world networks, respectively. Finally, we conclude in Section~\ref{sec:conclusion} by summarizing our work, and explaining how it can be extended.

\section{Related Work}
\label{subsec:relatedworks}
Authors historically interpreted the communities they found in an \textit{ad hoc} way \cite{Girvan2002,Rosvall2008,Blondel2008}, but this somewhat subjective approach does not scale well on large networks. 

More recently, several authors used topological measures to characterize community structures in plain networks. In \cite{Lancichinetti2010}, Lancichinetti \textit{et al}. visually examined the distribution of some community-based topological measures, both at local and intermediary levels. Their goal was to understand the general shape of communities belonging to networks modeling various types of real-world systems. In \cite{Leskovec2008}, Leskovec \textit{et al}. proposed to study the community structure as a whole, by considering it at various scales, thanks to a global measure called conductance. These two studies are valuable, however, from the interpretation perspective, they are limited by the fact they consider the network as a whole. Communities are studied and characterized collectively, in order to identify trends in the whole network, or even a collection of networks.

In order to characterize each community individually, some authors took advantage of the information conveyed by nodal attributes, when they are available. In \cite{Tumminello2011}, Tumminello \textit{et al}. proposed a statistical method to characterize the communities in terms of the over-expressed attributes found in the elements of the community. In \cite{Labatut2012}, Labatut \& Balasque interpreted the communities of a social attributed network. They used statistical regression and discriminant correspondence analysis to identify the most characteristic attributes of each community. Both studies are valuable, however they do not take advantage of the available topological measures to enhance the interpretation process.

Certain community detection methods take advantage of both relational (structure) and individual (attributes) information to detect communities. It seems natural to suppose the results they produce can be used for interpretation purposes. For example, in \cite{Zhou2009}, Zhou \textit{et al}. interpreted the communities in terms of the attributes used during the detection process; and in \cite{Yang2013}, Yang \textit{et al}. identified the top attributes for each identified community. However, the problem with these community detection-based methods is that the notion of community is often defined procedurally, i.e. simply as the output of the detection method, without any further formalization. It is consequently not clear how structure and attributes affect the detection, and hence the interpretation process. All these methods additionally rely on the implicit assumption of community homophily. In other words, communities are supposed to be groups of nodes both densely interconnected and similar in terms of attributes. To our knowledge, no study has ever shown this feature was present in all systems, or even in all the communities of a given network, or that all attributes were concerned. It is therefore doubtful those methods are general enough to be applied to any type of network.

Another method was recently defined based on frequent pattern mining, which can be used for community interpretation. In \cite{Stattner2012}, Stattner \& Collard introduced the notion of \textit{frequent conceptual link}. A conceptual link corresponds to a set of links from the original network, connecting nodes who share similar attributes. Such a link is said to be frequent when the number of links it represents is above a given threshold. This method can be seen as a generalization of the notion of homophily, and was initially used to simplify the network and help understanding it. Finding frequent conceptual links amounts to detecting groups of nodes sharing common attributes, with a pattern mining point of view. This methods considers both the network structure and the nodal attributes, however it ignores their evolution, i.e. it does not take the temporal aspect into account.

\section{Definitions \& Problem Statement}
\label{sub:Definition}
In this section, we first introduce some concepts used in the rest of the article. We then state formally the problem of community interpretation.

\subsection{Preliminary Definitions}
We first define several concepts related to community structures and complex networks in general. In particular, we describe a selection of topological measures later used in the experimental sections. In the second part, we focus on concepts related to sequential pattern mining.

\subsubsection{Network-Related Concepts}
\label{sec:interpreting}
We formally define a \textit{dynamic attributed network} as $G=\langle G_1,\ldots,G_\theta \rangle$, i.e. a sequence of chronologically ordered graphs $G_t$ ($1\leq t\leq\theta$), which we call time slices. A \textit{time slice} $G_t=(V,E_t,A)$ is a triple such that $V$ is the set of nodes, $ E_t \subseteq V \times V $ is the set of links and $A$ is the set of node attributes. The nodes are the same for all time slices, and we can consequently note $\mid V\mid=n$ the size of each time slice, as well as of the dynamic network. The set of nodal attributes is the same for all time slices, but the values associated to the nodes can change.

An \textit{evolving community structure} of a dynamic attributed network $G$ is a sequence $\langle \mathcal{C}_1,\cdots, \mathcal{C}_\theta\rangle$ of chronologically ordered community structures $\mathcal{C}_t$ ($1\leq t \leq \theta$), where each $\mathcal{C}_t$ corresponds to a community structure of $ G_t$. A \textit{community structure} $\mathcal{C}_t=\{C_t^1,...,C_t^{\lambda_t}\}$ is itself a partition of the node set, whose parts $C_t^c$ ($1\leq c \leq \lambda_t$) are the communities. We note $C_t (v)$ the function associating a node $v$ to its community in $\mathcal{C}_t$. The \textit{size} of a given community $ C_t^c $ is its number of nodes $ |C_t^c| $.
It is important to note that, due to the fact communities might split, merge or disappear, communities represented by the same index $c$ in two different time slices do not necessarily match. For instance, communities $C_4^1$ (first community at $t=4$) and $C_{10}^1$ (first community at $t=10$) might be completely different. 

A \textit{topological measure} quantifies the structural properties of the network or its components. Here, we focus on nine nodal measures, presented in the rest of this section. Each one will be processed for each node, at each time slice.

First, the \textit{degree} of a node is the number of links attached to it. More formally, we note $ N_{t}(v)=\{w\in V:\{v,w\}\in E_{t} \} $ the \textit{neighborhood} of node $ v $ at time $t$, i.e. the set of nodes connected to $ v $ in $ G_{t} $. The degree $ d_{t}(v)=|N_{t}(v)| $ of a node is the cardinality of its neighborhood, i.e. its number of neighbors at time slice $t$. Similarly, we can define the \textit{internal neighborhood} of a node $ v $ as the subset of its neighborhood located in its community: $ N_{t}^{int}(v)=N(v) \cap C(v) $. Then, the \textit{internal degree} $ d_{t}^{int}(v)=|N_{t}^{int}(v)| $ is defined as the cardinality of the internal neighborhood, i.e. the number of neighbors the node $ v $ has in its community.

The \textit{local transitivity} \cite{Duncan1998} corresponds to the ratio of existing to possible triangles containing $v$ in $G_t$:
\begin{equation}
  T_{t}(v)=\dfrac{|\{\{w_1,w_2\}\in E_{t}:w_1 \in N_{t}(v)\wedge w_2\in N_{t}(v)\}|}{d_{t}(v)(d_{t}(v)-1)/2} 
  \label{eqn:transitivity}
\end{equation}
\noindent In this ratio, the numerator corresponds to the observed number of links between the neighbors of $v$, whereas the denominator is the maximum possible number of such links. 

The \textit{eccentricity} of a node is its furthest distance to any other node in the network at a given time slice \cite{Harary1969}:
\begin{equation}
ecc_{t}(v)=\max_{w \in V}(dist_{t}(v,w))
\label{eqn:eccentricity} 
\end{equation}
\noindent Here, $ dist_{t}(v,w) $ is the geodesic distance between nodes $v$ and $w$ at time slice $t$. The geodesic distance corresponds to the length of the shortest path between two nodes.

The \textit{betweenness centrality}  measures how much a node lies on the shortest paths connecting other nodes. It is a measure of accessibility \cite{Freeman1979}:
\begin{equation}
K^{b}_{t}(v)=\sum_{i < j}\frac{\sigma^{ij}_{t}(v)}{\sigma^{ij}_{t}}
\label{eqn:betweenness} 
\end{equation}
\noindent Where $ \sigma^{ij}_{t}$ is the total number of shortest paths from node $ i $ to node $ j $, and $ \sigma^{ij}_{t}(v)$ is the number of shortest paths from $ i $ to $ j $ running through node $ v $ at time slice $t$.

The \textit{closeness centrality} quantifies how near a node is to the rest of the network, also in terms of geodesic distance \cite{Sabidussi1966}: 
\begin{equation}
K^{c}_{t}(v)=\frac{1}{\sum_{w \in V} dist_{t}(v,w)}
\label{eqn:closeness} 
\end{equation}

The \textit{Eigenvector centrality} measures the influence of a node in the network based on its spectral properties. The Eigenvector centrality of each node is proportional to the sum of the centrality of its neighbors \cite{Bonacich1987}: 
\begin{equation}
K^{e}_{t}(v)=\frac{1}{\lambda}\sum_{w \in N_{t}(v)}K^{e}_{t}(w)
\label{eqn:eigenvector} 
\end{equation}
\noindent Here, $\lambda$ is the largest Eigenvalue of the graph adjacency matrix.

The \textit{within module degree} and \textit{participation coefficient} are two measures proposed by Guimer\`{a} \& Amaral \cite{Guimera2005} to characterize the community role of nodes. The within module degree is defined as the $z$-score of the internal degree: 
\begin{equation}
z_{t}(v)= \frac{ d^{int}_{t}(v)-\mu(d^{int}_{t},C_{t}(v))}{\sigma(d^{int}_{t},C_{t}(v))} 
\label{eqn:withinmoduledegree} 
\end{equation}
\noindent Where $\mu$ and $\sigma$ denote the mean and standard deviation of $d^{int}_{t}$ over all nodes belonging to the community of $v$ at time slice $t$, respectively. This measure expresses how much a node is connected to other nodes in its community, relatively to this community. 

The \textit{participation coefficient} is based on the notion of community degree $ d^{c}_{t}(v)=|N_{t}(v)\cap C^{c}_{t}| $, which represents the number of links a node $ v $ has with nodes belonging to community $ C^{c}_{t} $:
\begin{equation}
P_{t}(v)= 1-\sum_{1\leq c \leq \lambda_{t}} (\frac{d^{c}_{t}(v)}{d_{t}(v)})^{2}
\label{eqn:participationcoeff} 
\end{equation}
\noindent Where $ \lambda_{t} $ is the number of communities in $ G_{t} $. $ P_{t} $ characterizes the distribution of the neighbors of a node over the community structure. More precisely, it measures the heterogeneity of this distribution: it gets close to $ 1 $ if all the neighbors are uniformly distributed among all the communities, and $ 0 $ if they are all gathered in the same community. 

The \textit{embeddedness} represents the proportion of neighbors of a node belonging to its own community \cite{Lancichinetti2010}. Unlike the within module degree, the embeddedness is normalized with respect to the node, and not the community:
\begin{equation}
e_{t}(v)= \frac{ d_{t}^{int}(v)}{d_{t}(v)} 
\label{eqn:embeddedness} 
\end{equation}

\subsubsection{Pattern-Related Concepts}
\label{subsubseq:PatternRelatedConcepts}
A \textit{node descriptor} is either a topological measure or a node attribute from $A$. Let $D=\{D_1,D_2,\cdots,D_k \}$ be the set of all descriptors. Each descriptor from $D$ can take one of several discrete values, defined in its domain $\mathcal{D}_{i}$ ($1\leq i\leq k$). All our topological measures are real-valued, so we have to discretize them to fit this definition. Moreover, the same apply to real-valued attributes. The details of the discretization and binning processes are explained in Section~\ref{subsub:dataprep}.

An \textit{item} $l_i=(D_i,x)\in D\times \mathcal{D}_{i} $ is a couple constituted of a descriptor $D_i$ and a value $x$ from its domain $\mathcal{D}_{i}$. The set of all items is noted $I$. An \textit{itemset} $h$ is any subset of $I$. Although itemsets are sets, in the rest of this article we represent them between parentheses, e.g. $h=(l_1,l_3,l_4 )$ because it is the standard notation in the literature. 

A \textit{sequence} $s= \langle h_1,\cdots , h_m \rangle $ is a chronologically ordered list of itemsets. Two itemsets can be consecutive in the sequence while not correspond to consecutive time slices: the important point is that the first to appear must be associated to a time slice preceding that of the second one. In other words, $h_i$ occurs before $h_{i+1}$ and after $h_{i-1} $. The \textit{size} of a sequence is the number of itemsets it contains. 

A sequence $\alpha=\langle a_{1},\ldots,a_{\mu} \rangle$ is a \textit{sub-sequence} of another sequence $\beta=\langle b_{1},\ldots,b_{\nu}\rangle$ iff $\exists i_{1},i_{2},\ldots,i_{\mu}$ such that $1\leq i_{1}<i_{2}<\ldots<i_{\mu}\leq \nu$ and $a_{1}\subseteq b_{i_1},a_{2}\subseteq b_{i_2}, \ldots,a_{\mu}\subseteq b_{i_{\mu}}$. This is noted $\alpha\sqsubseteq\beta$. It is also said that $\beta$ is a \textit{super-sequence} of $\alpha$,which is noted $\beta\sqsupseteq\alpha$.

The \textit{node sequence} $u(v)$ of a node $v$ is a specific type of sequence of size $\theta$ (i.e. the number of time slices). We have $u(v)=\langle(l_{11},\cdots,l_{k1} )\cdots (l_{1\theta},\cdots,l_{k\theta} ) \rangle$, where $l_{it}$ is the item containing the value of descriptor $D_i$ for $v$ at time $t$. A node sequence $u(v)$ includes $\theta$ itemsets, i.e. it represents all time slices. Each one of these itemsets contains all $k$ descriptor values for the considered node at the considered time. In other words, $u(v)$ contains all the available descriptor-related data for node $v$. 

We build the \textit{enlarged node sequence} $u_{enl}(v)$ of a node $v$ by adding a community-related item to each itemset of its node sequence: $ u_{enl}(v)=\langle(l_{11},\cdots,l_{k1}, C_1(v))\cdots(l_{1\theta},\cdots,l_{k\theta},C_{\theta}(v)) \rangle $. The \textit{sequence database} $M$ can then be obtained by collecting the enlarged node sequences $u_{enl}(v)$ of all nodes in the considered network.

The set of \textit{supporting nodes} $S(s)$ of a sequence $s$ is defined as $S(s) =\{v\in V:u(v) \sqsupseteq s\} $. The support of a sequence $s$, $Sup(s)= |S(s)|/n$ , is the proportion of nodes, in $G$, whose node sequences are equal to $s$, or are super-sequences of $s$. 

The set of \textit{supporting nodes} of a sequence $s$ \textit{relatively} to a node group $X$ is defined as $S(s,X )=\{v\in X:u(v)\sqsupseteq s\}$. Its support relatively to the same node group, $Sup(s,X)= |S(s,X)|/|X|$, is the proportion of nodes, in $X$, whose node sequences are equal to $s$, or super-sequence of $s$. A node group might directly correspond to a community taken at one time slice, or to the nodes belonging to the same communities over a series of time slices.

The \textit{growth rate} of a pattern $s$ relatively to a node group $X$ is $Gr(s,X)=Sup(s,X)/Sup(s,\overline{X}) $, where $\overline{X}$ is the complement of $X$ in $V$, i.e. $ X=V\setminus\overline{X} $. The growth rate measures the \textit{emergence} of $s$: a value larger than $1$ means $s$ is particularly frequent (i.e. emerging) in $X$, when compared to the rest of the network.

We say a sequence is \textit{community-related} if it contains at least one community-related item. If all its items are community-related, it is said to be a \textit{community sequence}, such as $ \langle C_{t_1}^{c_1},C_{t_2}^{c_2},\cdots,C_{t_m}^{c_m} \rangle $. On the contrary, we call it \textit{community-independent} if it contains no community-related item at all. 

For a community-related sequence $s$, we define its \textit{community-wise sub-sequence} $s_{wise}$ as its maximal community sub-sequence. In other words, it is a sequence $s_{wise} \sqsubseteq s$ such that $s_{wise}$ is a community sequence and there is no other community sequence $s'$ fulfilling both conditions $s' \sqsubseteq s$ and $s_{wise}\sqsubset s'$ . Similarly, for a community-related sequence $s$, we define its \textit{community-less sub-sequence} $s_{less}$ as its maximal community-independent sub-sequence. In other words, it is a sequence $ s_{less} \sqsubseteq s$ such that $s_{less}$ is a community-independent sequence and there is no other community-independent sequence $s'$ fulfilling both conditions $s' \sqsubseteq s$ and $s_{less}\sqsubset s'$. 

Given a minimum support threshold noted $min_{sup}$, a \textit{frequent sequential pattern} (FS) is a sequence whose support is greater or equal to $min_{sup}$. A \textit{closed frequent sequential pattern} (CFS) is a FS which has no super-sequence possessing the same support.

\subsection{Problem Statement}\label{subsec:problem}
We see the problem of \textit{community interpretation} as the operation consisting in identifying the most \textit{characteristic features} of certain groups of nodes. But \textit{what} is a feature? And \textit{how} can we know if it is a characteristic one? To be able to solve the interpretation problem, we need first to answer these two questions. In other words, our problem of interest can be broken down to two sub-problems: 
\begin{enumerate}
  \item \textit{Finding an appropriate way to represent a community;}
  \item \textit{Defining an objective method to decide which parts of this representation are characteristic. }
 \end{enumerate}
In this section, we consider separately these two sub-problems and formalize them.

\subsubsection{Appropriate Community Representation}
\label{subsubsec:CommunityRepresentation}
It is obviously not possible to know in advance which pieces of the information describing the considered community will be the most characteristic. Therefore, we need to be able to represent all the available information, in a computationally efficient way. A community can be described only in terms of its constituting elements, since it is by definition a set, so its representation must be defined at the nodal level. It is necessary to use a representation able to handle node similarity, interconnection, and co-evolution. More formally, this means a community must be represented through its nodal attributes, topological properties and temporal evolution. The attributes correspond to the individual characteristic of the objects composing the modeled social system, the topological properties describe how these objects interact, and the temporal evolution is the consequence of the system dynamics.

\begin{quote}
\textit{In the context of community interpretation, the problem of \textbf{finding an appropriate community representation}, for a dynamic attributed network and its community structure, is equivalent to that of encoding all information describing the evolution of each node from each community. This encoding should be compact enough to avoid redundancies, but complete enough to describe the evolution of the community.}
\end{quote}

To fulfill these constraints, we propose to represent each node using the \textit{sequence} of its attributes and topological measures, taken at different discrete times of the system evolution. To our knowledge, such a sequential representation was never used for networks before. We could use the database $M$ described in the previous section, which is the collection of enlarged node sequences $u_{enl} (v)$ ($\forall v\in V$). However, this representation would not be very compact, because several nodes could be described (totally or partially) by similar sequences. To avoid this, we instead represent a community through the sequential patterns present among the sequences describing the nodes it contains. 

For a node $v$, we note $P_v$ the set of all possible sub-sequences of its enlarged node sequence, i.e. $P_v=\{s\sqsubseteq u_{enl}(v) \}$. Let $P$ be the union of all the $P_v$ over all nodes, i.e. $P=\bigcup_{v\in V}P_v$, and $B$ be the subset of its community sequences. Let $m_i$ denote such a community sequence and $\mu$ be the cardinality of $B$, then we have $B=\{m_1,\cdots,m_{\mu} \}$. We state the problem of community representation is finding a set $\Gamma=\{\gamma_1,\cdots,\gamma_{\mu} \} $ such that $\gamma_{i}=\{s\in P:s\sqsupseteq m_i \}$ with $1 \leq i \leq \mu$. In other words, for each community sequence $m_i$ found in the database, we want to identify the set $\gamma_{i}$ of all its super-sequences present in the same database. Those, by definition, are themselves community-related sequences. 

The set $\Gamma$ includes all the necessary information related to each community sequence. It represents the sub-sequences common to more than one node only once, and therefore eliminates the redundancies. Consequently, it fulfills our requirements of being both a complete and compact representation.

\subsubsection{Identifying Characteristic Features}
\label{subsubsec:secondprob}
Let us now turn to the second problem: finding, in an objective way, which parts of the community description are characteristic. The criteria used to identify this relevance must be compatible with our representation of a community, which takes the form of sets of sequential patterns. 

\begin{quote}
\textit{In the context of community interpretation, the problem of\textbf{ identifying characteristic features} among the sequences representing a community consists in selecting some objective criteria to assess the representative power of these sequences, and a method to select the most representative ones.}
\end{quote}

The representation defined in the previous section takes the form of a set $\Gamma=\{\gamma_1,\cdots,\gamma_{\mu}\}$, where each $\gamma_i$ represents the set of community-related super-sequences of a community sequence $m_i$ (as of the considered database). We define one condition and two criteria for a member of $\gamma_i$ to be characteristic of $m_i$. The condition is that it must be \textit{informative}. We consider a sequence to be informative if it is closed \cite{Gallo2007}, i.e if it does not have any super-sequence with the same support, or a better one. The two criteria are that the sequence must be both \textit{prevalent} and \textit{distinctive}. We measure the prevalence of a sequence $s$ in $\gamma_i$ with its support $Sup(s,S(m_i ) )$, where $S(m_i)$ is the supporting node set of the reference community sequence $m_i$ (i.e. the groups of nodes following the sequence). We measure the distinctiveness through the growth rate $Gr(s,S(m_i))$ \cite{Dong1999}. 

Then, the problem of identifying the characteristic features of a community sequence $m_i$ consists in selecting a subset $\gamma_{i}^{'}\subseteq \gamma_i$ such that $\gamma_{i}^{'}=\{s\in \gamma_i: s\; is\; closed\wedge Sup(s,S(m_i ) )\geq min_{sup} \wedge Gr(s,S(m_i ) )\geq min_{gr}\}$, where $min_{sup}$ and $min_{gr}$ are lower thresholds for the support and growth rate, respectively. We note $\Gamma^{'}=\{\gamma_1^{'},\cdots,\gamma_{\mu}^{'} \}$ the set containing the characteristic patterns of each community sequence in the network.

\section{Overview of our Evolution-Based Approach}
\label{sec:overview}
\begin{figure*}
	\center
  \includegraphics[width=1.0\textwidth]{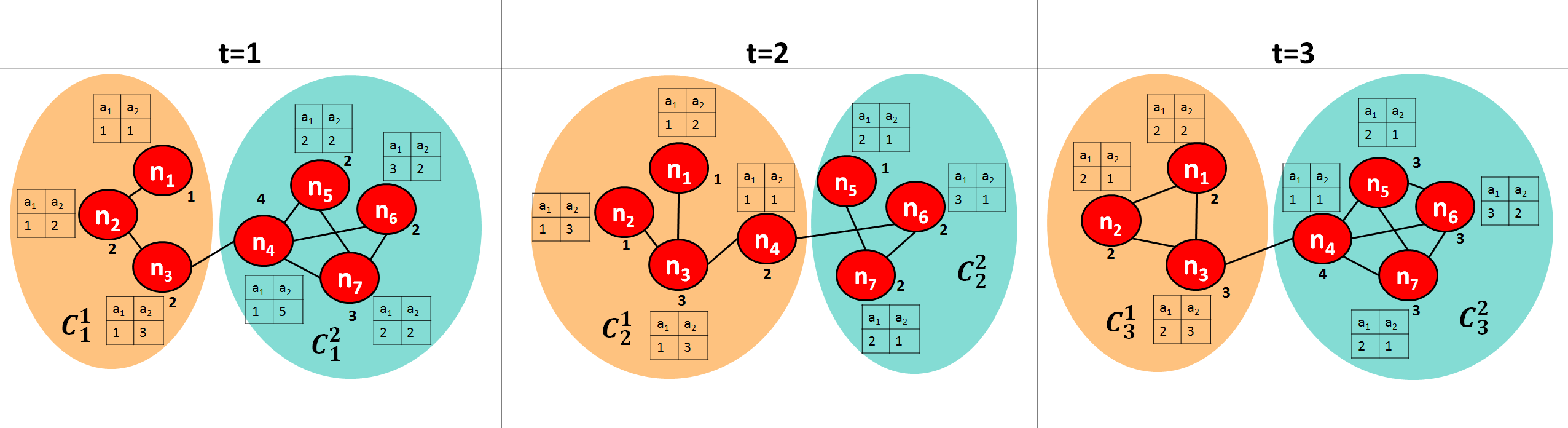}
\caption{Example of attributed dynamic network, including three time slices and two nodal attributes $a_1$ and $a_2$.}
\label{fig:toyNetwork}       
\end{figure*}

Our approach is based on the representation of dynamic attributed networks under the form of sequences. One sequence represents a node and its behavior: it contains the topological, attribute and community-related information describing it for each time slice of the studied period. Let us illustrate our approach on a small network of Jazz listeners, extracted from the data analyzed in Section \ref{subsec:LastFM}. This network includes $7$ nodes whose connections are changing over $3$ time slices. Each node corresponds to a user of the LastFM service, i.e. a listener, and each link represents a friendship relation between two users. We consider two nodal attributes: number of times a user listens to Miles Davis ($a_1$) and to Chet Baker ($a_2$) in the considered time slice. Regarding the topological information, we limit this example to the sole degree, which can be interpreted here as a measure of the users' popularity. The network contains two communities, which also change through time. Figure~\ref{fig:toyNetwork} represents the network itself, the nodal attributes and topological measure, as well as the communities.

The collection of sequences describing the network is shown in Table~\ref{tab:exampleTable}. For example, the sequence representing node $n_4$ is: $\langle (a_1=1,a_2=5,deg=4,C_{1}^{2}) (a_1=1,a_2=1,deg=2,C_{2}^{1})(a_1=1,a_2=1,deg=4,C_{3}^{2})\rangle$, where $C_t^c$ denotes the group of nodes corresponding to community number $c$ at time $t$. This sequence means the user $n_4$ listens to Miles Davis once and Chet Baker five times, has four friends and belongs to the second community during the first time slice. Then, he listens to Miles Davis and Chet Bakers once, loses two friends and switches to the first community during the second time slice. Finally, he listens to Miles Davis and Chet Bakers once again, gains two friends and switch back to the second community during the last time slice.

\begin{table}
\caption{Node sequences representing the network from Fig.~\ref{fig:toyNetwork}.}
\label{tab:exampleTable}       
  \begin{tabular}{|p{0.02\textwidth}|p{0.40\textwidth}|}
    \hline
 	\textbf{ID}  &  \textbf{Related sequence}\\  
     \hline
    \textbf{$n_1$}  &  
    $\langle (a_1=1,a_2=1,deg=1,C_{1}^{1}) (a_1=1,a_2=2,deg=1,C_{2}^{1})(a_1=2,a_2=2,deg=2,C_{3}^{1})\rangle$ \\  
 	\textbf{$n_2$}  &  
 	$\langle (a_1=1,a_2=2,deg=2,C_{1}^{1}) (a_1=1,a_2=3,deg=1,C_{2}^{1})(a_1=2,a_2=1,deg=2,C_{3}^{1})\rangle$ \\  
 	\textbf{$n_3$}  &  
 	$\langle (a_1=1,a_2=3,deg=2,C_{1}^{1}) (a_1=1,a_2=3,deg=3,C_{2}^{1})(a_1=2,a_2=3,deg=3,C_{3}^{1})\rangle$ \\  
 	\textbf{$n_4$}  &  
 	$\langle (a_1=1,a_2=5,deg=4,C_{1}^{2}) (a_1=1,a_2=1,deg=2,C_{2}^{1})(a_1=1,a_2=1,deg=4,C_{3}^{2})\rangle$\\  
 	\textbf{$n_5$}  &  
 	$\langle (a_1=2,a_2=2,deg=2,C_{1}^{2}) (a_1=2,a_2=1,deg=1,C_{2}^{2})(a_1=2,a_2=1,deg=3,C_{3}^{2})\rangle$\\  
 	\textbf{$n_6$}  &  
 	$\langle (a_1=3,a_2=2,deg=2,C_{1}^{2}) (a_1=3,a_2=1,deg=2,C_{2}^{2})(a_1=3,a_2=2,deg=3,C_{3}^{2})\rangle$ \\  
 	\textbf{$n_7$}  &  
 	$\langle (a_1=2,a_2=2,deg=3,C_{1}^{2}) (a_1=2,a_2=1,deg=2,C_{2}^{2})(a_1=2,a_2=1,deg=3,C_{3}^{2})\rangle$\\  
 	\hline
   \end{tabular}
\end{table}

Based on this representation, we search for the most characteristic sequences related to communities. More specifically, we look for closed frequent emerging sequences, as explained in details in Section~\ref{sec:method}. Because our focus is on communities, we want to uncover only sequences containing community-related items. For example, the simple sequence $s_1=\langle (deg=2)\rangle$ is very frequent, since every node has two friends at least once during the time period. However, $s_1$ does not include any information about the communities, so we ignore it. Sequence $s_2=\langle (a_1=1,C_{1}^{1}) (a_1=1,C_{2}^{1})(a_1=2,C_{3}^{1})\rangle$, on the contrary, is a bit less frequent (supported by the first $3$ nodes), but contains community information. It can be interpreted as a trend regarding the listening habits of certain nodes of the first community.

At this point, it is important to understand a community is not necessarily a stable group of nodes. On the contrary, it can undergo rather dramatic changes during its evolution: merge, split, complete disappearance, etc. To handle this case, it is necessary to focus the interpretation process on groups of nodes going through several times slices together, \textit{while possibly switching communities simultaneously}. For this purpose, we separate sequences such as $s_2$ in two subsequences: on the one hand, a community sequence $\langle (C_{1}^{1})(C_{2}^{1})(C_{3}^{1})\rangle$ containing exclusively community-related items, and used to represent how the concerned nodes evolve community-wise ; and on the other hand, a characteristic sequence $\langle (a_1=1) (a_1=1)(a_1=2)\rangle$ containing no community-related items at all, which serves as a basis to interpret this group of nodes. This is better illustrated when considering the second community: the community pattern $\langle (C_{1}^{2})(C_{3}^{2})\rangle$ allows covering nodes $n_4$ to $n_7$, whereas adding $(C_{2}^{2})$ would exclude $n_4$. 

Our last point concerns the covering of the community sequences. For $\langle (C_{1}^{1})(C_{2}^{1})(C_{3}^{1})\rangle$, we identified $s_2$, which describes all $3$ concerned nodes. However, it is not always possible to do so: it is noticeably the case for $\langle (C_{1}^{2})(C_{3}^{2})\rangle$. For this community sequence, the best characteristic sequence is $s_3=\langle (a_1=2)(a_1=2, a_2=1)(a_1=2,a_2=1,deg=3)\rangle$, which is supported only by $n_5$ and $n_7$, but not $n_4$ and $n_6$. In this situation, we identify supplementary sequences such as $s_4=\langle (a_2=1,deg=2)\rangle$ to improve the coverage, and use them to complement the interpretation.

\section{Proposed Method}
\label{sec:method}
Our problem definition requires us to search for informative, prevalent and distinctive sequences by taking advantage of a compact and complete representation of dynamic attributed networks. To fulfill this goal, we propose a two-stepped approach. In the first, we build a sequence database, in order to represent a dynamic attributed network. The second then consists in searching for meaningful sequences while respecting the constraints defined in Section~\ref{subsubsec:secondprob}. 

\subsection{Creating the Sequence Database}
\label{subsec:createdb}
As explained in Section ~\ref{sec:interpreting}, a sequence database is the collection of enlarged node sequences $u_{enl}(v)$ for all nodes in the network. Creating this database thus requires first calculating the topological measures and identifying the communities. Moreover, all real-valued nodal descriptors must be discretized to allow sequential pattern mining. Data preparation may also include some additional preprocessing, such as the binning of discrete descriptors or the combination of topological measures, in order to improve the readability of the obtained patterns, or to lighten the computational load.

\subsubsection{Community Detection}
\label{subsubsec:CommunityDetection}
An important idea with our approach is to interpret the communities independently from the method used for their detection. But of course, we need a reference community structure to work with. Community detection for time evolving networks is much less developed than for static networks. Moreover, it was shown in \cite{Aynaud2010} that applying static community detection methods on evolving networks does not lead to stable results, which shows there are non-negligible differences between these two problems. Note that our goal here is not to perform an exhaustive comparison of algorithms able to process dynamic networks: the focus of this article is not community detection itself, but rather the interpretation of the detected communities.

In \cite{orman2014contribution}, the author tested four different versions of the Louvain algorithm \cite{Blondel2008}, modified for dynamic networks. She has shown that \textit{Incremental Louvain} \cite{Aynaud2010} was generally above the others in terms of performance. So, on the basis of this study, we selected this algorithm to detect evolving communities in our own data. Unlike other variants of Louvain, Incremental Louvain takes into account the previous communities when processing a time slice, which results in some temporal smoothing. For the first time slice, the original version of Louvain is applied. Then, for the next time slices, the algorithm starts with the community structure found at the previous time slice (rather than putting each node in its own community), and then goes on with the original processing (greedy optimization of the modularity measure, see \cite{Blondel2008}). The time complexity of this method is in $O(n \log n )$ \cite{Blondel2011} for one time slice when the network is sparse. In our case, it is applied on $\theta$ time slices, so the total complexity of is in $O( \theta n \log n)$.

\subsubsection{Data Preparation}
\label{subsub:dataprep}
As explained in Section \ref{subsubseq:PatternRelatedConcepts}, the descriptors used to fill our sequence database are either nodal attributes or topological measures. All the measures we selected are numerical, whereas the attributes can be of any data type, including categorical and numerical values. However, it is not possible to directly use real values as items for sequential pattern mining, because this method handles only discrete data. So, it is necessary to first \textit{discretize} this type of descriptor. Moreover, even for integer descriptors, it can be interesting to \textit{bin} their domains for several reasons. First, having too many distinct values in the domain of a descriptor tends to significantly increase the number of detected patterns while decreasing their support, thereby preventing to uncover results which would be sufficiently general to be informative. Second, due to the increase in the number of patterns, both processing time and memory occupation also significantly increase during pattern mining (as shown in Section \ref{sec:Eval}). Three, the meanings of two distinct descriptor values can be close enough that they can be considered as similar without any significant information loss. 

Whatever the preparation method (discretizing or binning), it is necessary to define thresholds. This can be done either by taking advantage of some expertise regarding the system or measures, or by using an automatic approach, for instance through the identification of denser zones in the considered attribute domain. A clustering algorithm can fulfill this goal, with the added advantage of being able to handle simultaneously \textit{several} descriptors. Indeed, for computational reasons, as well as to ease human interpretation, it can be relevant to go beyond discretization and binning, and to \textit{combine} several descriptors into one. This approach was previously applied, in the context of social networks analysis, to some variants of Guimera \& Amaral's measures, in order to identify community roles \cite{Dugue2014}.

We propose to generalize this idea to the preparation of all the topological measures we use in this work (Section \ref{sec:interpreting}). We first distinguish three groups of thematically related measures: centrality-related (eccentricity, betweenness, closeness, Eigenvector), community-related (embeddedness, within module degree, participation coefficient), and local (degree, local transitivity). The first group concerns the position of the node in the whole network, the second focuses on the community structure, and the third represents its local connectivity. Each group is clustered separately using the $k$-means algorithm, which was chosen because it is a well-known and fast method. The number of clusters (parameter $k$) is decided by optimizing the average Silhouette width \cite{Rousseeuw1987}, a widespread cluster quality measure whose interpretation is clearly defined. After this process, our set of $9$ topological measures is replaced by $3$ discrete descriptors, whose values correspond to the detected clusters.

\subsection{Mining the Sequence Database}
\label{subsec:minedb}
The first step of our mining process consists in identifying all CFS, relatively to the support threshold $min_{sup}$. Among them, the most distinctive ones are selected using the growth rate threshold $min_{gr}$. Finally, an additional filtering is performed to only keep the sequences necessary for a good coverage of the considered communities. This section is dedicated to the description of these three steps.

\subsubsection{Mining the Closed Frequent Sequences}
We use \textit{CloSpan} (\underline{Cl}osed \underline{S}equential \underline{Pa}tter\underline{n} Mining) \cite{Yan2003} to mine the CFS relatively to our $min_{sup}$ threshold. It is an efficient algorithm able to identify long sequences in real-world data, in a practical time. It relies on the mining strategy introduced in \textit{PrefixSpan} \cite{Pei2001}.

At first, CloSpan creates a candidate set, which is a super-set of the closed frequent sequences. These candidates are stored into a so-called \textit{prefix sequence lattice}. Second, the non-closed sequences are eliminated. A na\"{\i}ve approach consists in checking, for all candidate sequences, if there is any super-sequence with the same support in the prefix sequence lattice. But it is a costly operation. Thus, Yan \textit{et al}. adopted the fast subsumption checking algorithm introduced in \cite{Zaki2002}. It is designed to manage a hash table in which, for each sequence $s$, the associated hash key is the sum of the corresponding sequences IDs. Here, the corresponding sequences of a sequence $s$ refer to all sequences with prefix $s$, i.e. the members of its \textit{projected database} (c.f. \cite{Yan2003} for the explanation and formalization of this notion of \textit{projected database} of a sequence). In \cite{Li2006}, the authors claim the time complexity of the last step of CloSpan (pruning prefix sequence lattice, the most demanding step) is in $O(n^2)$, where $n$ is the size of the data (in our case: the number of nodes), if the maximum length of the frequent sequences is constrained by a constant.

\subsubsection{Identifying the Emerging Patterns}
The emergence is assessed by processing the growth rates of the sequences, as described in Algorithm~\ref{code:enlargedGr}. CloSpan outputs all the CFS related to the considered database, as well as their support, and these data constitute the input of our algorithm, together with the database $M$ itself. Each sequence is processed separately. 

\begin{algorithm}
\caption{Identification of the Characteristic Sequences}
\label{code:enlargedGr}
\begin{algorithmic}[1] 
\REQUIRE {$CFS$, $M$, $Sup[~]$}
\ENSURE{$B=\{m_1,...,m_{\mu}\}$, $\Gamma'=\{\gamma_{1}^{'},...,\gamma_{\mu}^{'}\}$, $Sup[~]$, $Gr[~]$}

\STATE  {$i \leftarrow 0$}
\FOR{$all$ $s\in CFS$ }
	\IF{isCommunityRelated($s$)}
		\STATE  {$i \leftarrow i + 1$}
		\STATE  {$(s_{less},\; s_{wise} ) \leftarrow$ separate($s$)}
		\STATE  {$m_i \leftarrow s_{wise}$}
		\STATE  {$Sup[ s_{wise} ] \leftarrow$ processSup($s_{wise}$, $M$)}
		\STATE  {$Sup[ s_{less} ] \leftarrow$ processSup($s_{less}$, $M$)}
		\STATE  {$Gr[ s_{less}, S(s_{wise})] \leftarrow$ processGr($Sup[~]$, $s_{wise}$, $s_{less}$, $n$)}
		\IF{$Gr[ s_{less}, S(s_{wise})] \geq min_{gr} $}
			\STATE  {$\gamma_{i}^{'} \leftarrow \gamma_{i}^{'} \cup \{s_{less}\}$}
		\ENDIF
	\ENDIF
\ENDFOR
\end{algorithmic}
\end{algorithm}

The first step consists in determining if the sequence is relevant. Indeed, CloSpan identifies both the sequences showing general trends over the whole network (i.e. community-independent CFS) and those relative to community sequences (i.e. community-related CFS). In our situation, we need to focus only on the latter, so it is necessary to first separate them from the former. This is done through the function \texttt{isCommunityRelated}, whose complexity is in $O(\theta)$, where $\theta$ is the size of the longest possible sequence.

For each remaining CFS $s$, we apply a parsing procedure in order to break it down to its \textit{community-wise} and \textit{community-less} sub-sequences, noted $s_{wise}$ and $s_{less}$, respectively. The former is simply the community sequence of interest, whereas the latter is potentially one of its characteristic sequences. This task is performed by the function \texttt{separate}, which is also in $O(\theta)$. Then, $s_{wise}$ is added to $B$, which gathers all community sequences. We remind the reader that, according to the definitions from Section~\ref{subsubsec:CommunityRepresentation}, $B=\{m_1,...,m_{\mu}\}$ is meant to eventually contain all community sequences $m_i$.

Next, we want to process the growth rate of $s_{less}$ for the supporting nodes of $s_{wise}$. To this aim, we need to retrieve the supports of both these sub-sequences. We use the function \texttt{processSup} to process the growth rate of a given sequence $s$. It first looks $s$ up in the CFS outputted by CloSpan: if the sequence is closed, then its support is directly available. Otherwise, it must be processed. This requires considering each node sequence from the database $M$ ($O(n)$ operations in the worst case), and checking if it is a super-sequence of $s$ ($O(\theta^2)$ operations in the worst case). The total complexity of the function is therefore in $O(\theta^2 n)$.

Using the supports, the growth rate can be processed in constant time thanks to the function \texttt{processGr}. It can then be used to discard non-emerging sequences according to our $min_{gr}$ threshold. On the contrary, emerging sequences are added to $\Gamma'$. We remind the reader that, as defined in Section~\ref{subsubsec:secondprob}, $\Gamma'=\{\gamma_{1}^{'},...,\gamma_{\mu}^{'}\}$ is meant to eventually contain all sets $\gamma_{i}^{'}$, each one gathering all the characteristic sequences associated to community sequence $m_i \in B$.

The complexity of the operations contained in the \texttt{For} loop is in $O(\theta^2 n)$. If we assume CloSpan outputted $r$ CFS, the total complexity of the whole algorithm is thus in $O(r \theta^{2} n)$.

\subsubsection{Selecting the Characteristic Patterns}
The main output of the previous step is $\Gamma^{'}=\{\gamma_1^{'},\cdots,\gamma_{\mu}^{'} \}$. Each $\gamma_i^{'}$ contains community-independent sequences associated with the community sequence $m_i$. By construction, all the sequences in $\Gamma'$ are closed, frequent and emerging, and we therefore consider them as characteristic. However, there can still remain too many of them to perform a relevant interpretation. As a post-process, we propose an additional filtering, leading to smaller sets noted $\Gamma^{''}=\{\gamma_1^{''},\cdots,\gamma_{\mu}^{''} \}$, and such that $\gamma_{i}^{''}\subseteq \gamma_{i}^{'} $.

We can either consider the sequences with highest growth rate, or with highest support. The complementary selection procedure we propose is generic, and can be applied to both cases. In the rest of our explanations, we refer to the criterion of interest (growth rate or support) as the sequence \textit{score}. Once the sequence with highest score has been selected, there is no guarantee for it to cover a sufficient part of the studied community sequence. And indeed, in practice it appears to be the opposite (especially for the growth rate). It is thus needed to identify other complementary sequences, allowing us to obtain a more complete coverage of the nodes supporting the community sequence. 

Intuitively, we want to find a small number of patterns, such that they cover a significant part of the community sequence, and are different in terms of supporting nodes. Or, more formally:
\begin{itemize}
	\item The cardinality of $\bigcap_{s \in \gamma_{i}^{''}} S(s,m_i ) $ must be minimal;
	\item The cardinality of $\bigcup_{s \in \gamma_{i}^{''}} S(s,m_i ) $ must be maximal (if possible: the whole community);
	\item The cardinality of $\gamma_{i}^{''}$ must be minimal. 
\end{itemize}

In order to perform this selection, we apply an iterative procedure described by Algorithm~\ref{code:patternselection}. We treat each detected community sequence $m_i$ separately. The sequence set \texttt{Remaining} represents the characteristic sequences of $m_i$ not treated yet, which is why it is initialized with $\gamma'_i$. The node set \texttt{Covered} contains the nodes supporting $m_i$ which are currently also supporting at least one sequence in $\gamma''_i$, whereas \texttt{Uncovered} contains those who are not. The following processing is then iterated until those sets stabilize. First, we apply the function \texttt{choose}, which is designed to return a sequence $s$ from $\gamma''_i$ which was not used yet, and is optimal for our criteria. Its complexity is in $O(n)$ (number of nodes). We then use $s$ to update $\gamma''_i$ and the three working sets. At the end of the \texttt{Repeat} loop, the nodes still uncovered are considered as anomalies.

\begin{algorithm}
\caption{Additional Filtering of the Sequences}
\label{code:patternselection}
\begin{algorithmic}[1]
\REQUIRE {$B$, $\Gamma^{'}$, $Score[~]$, $S[~]$, $max_{seq}$  }
\ENSURE{$\Gamma''=\{\gamma_{1}^{''},...,\gamma_{\mu}^{''}\}$ }


\FOR {$i$ in $[1,\mu]$}
	\STATE  {$Remaining\leftarrow \gamma_{i}^{'}$}
	\STATE  {$Uncovered\leftarrow S(m_{i}) $ }
	\STATE  {$Covered\leftarrow \varnothing  $ }

	\REPEAT
		\STATE  {$s \leftarrow$ choose($Remaining$, $Uncovered$, $Covered$, $Score$)}
		\STATE  {$\gamma_{i}^{''} \leftarrow \gamma_{i}^{''} \cup \{s\}$}
		\STATE  {$Remaining \leftarrow Remaining \setminus \{s\}$}
		\STATE  {$Uncovered \leftarrow Uncovered \setminus S(s)$}
		\STATE  {$Covered \leftarrow Covered \cap S(s)$)}
	\UNTIL {$Uncovered$ does not change $\vee$ $|\gamma_{i}^{''}| \geq max_{seq}$ }
\ENDFOR
\end{algorithmic}
\end{algorithm}

The \texttt{Repeat} loop is processed at most $max_{seq}$ times, where $max_{seq}$ is a parameter defined by the user. Indeed, the goal of this post-processing is to reduce the number of selected characteristic patterns, so it is necessary to set a limit corresponding to a subjective acceptable number. The \texttt{For} loop is repeated exactly $\mu$ times, so the total complexity of this algorithm is in $O(\mu n)$. If we consider the two first steps of our mining method, we get a final complexity of $O(n^2 + r \theta^{2} n + \mu n)$. In practice, CloSpan detects many more CFS than there are nodes in the network, so $r \gg n$, and therefore $rn \gg n^2$. The number of community sequences $\mu$ is bounded by the total number of sequences $r$, so we can neglect the last term. We finally obtain the following simplified expression: $O(r \theta^{2} n)$. In the end, the complexity of our tool depends essentially on the number of nodes ($n$) and time slices ($\theta$) in the studied network, and the number of sequences outputted by CloSpan ($r$).

\section{Evaluation on Artificial Networks}
\label{sec:Eval}
In this section, we describe the experiments carried out to study how changes in the data affect the performance of our method. To control these changes, we relied on some artificially generated datasets. We first describe the model used to generate the data, then present our experimental results.

\subsection{Generative Model}
The level of realism of the generated networks is known to have an effect on certain analysis tools, in particular community detection algorithms \cite{Lancichinetti2008,Orman2010}. To the best of our knowledge, the model generating community-structured networks with the most realistic topology is LFR \cite{Lancichinetti2008}. 
However, it was designed to produce \textit{static} networks. Recently, it was extended to generate \textit{dynamic} networks with predefined evolving community structures \cite{Greene2010}. We call this model LFR-D. Unfortunately, this extension does not handle nodal attributes. This is why we propose to further extend it, leading to the LFR-DA model, able to also generate nodal attributes. We first describe briefly the LFR and LFR-D models, before introducing our own extension LFR-DA.

\subsubsection{Generating Time Evolving Networks}
The LFR model of Lancichinetti \textit{et al}. \cite{Lancichinetti2008} first uses the Configuration Model \cite{Molloy1995} to generate a network without any community structure, but whose size and degree distribution are controlled. A rewiring process then takes place to make communities appear while preserving the degree distribution. This leads to a static network.

The LFR-D model of Greene \textit{et al}. starts with a static community-structured network outputted by LFR. This network is used as the first time slice, and its structure is then altered to produce the following time slices through the occurring of community-related events. The user specifies the number of desired time slices. There are $5$ types of community events, separated in two classes. On the one hand, \textit{large-scale} events: \textit{birth-death}, \textit{merge-split}, \textit{hide-appear}; and on the other hand, small-scale events: \textit{expansion-contraction} and \textit{switch}. 

Large-scale events cause dramatic changes in the community structure. They include the creation of a new community (\textit{birth}), the deletion of an existing one (\textit{death}), the separation of an existing community into several smaller new ones (\textit{split}), the union of several communities into a larger new one (\textit{merge}), and the temporary disappearance of a community (\textit{hide} then \textit{appear}). These events are controlled by user-defined parameters determining the number of communities to be modified at each time slice. For example, if the event type is birth-death, LFR-D takes two parameters: the numbers of communities to be created and to be removed at each time slice.

Small-scale events correspond to local modifications, and are not likely to cause important changes in the community structure. They include the growing or shrinking of an existing community (\textit{expansion} and \textit{contraction}), and the shift of a few nodes from one community to another (\textit{switch}). For these events, the user specifies a parameter corresponding to the proportions of nodes concerned by the modification. For expansion-contraction, it also takes the numbers of communities whose size must be increased or decreased, respectively, at each time slice.

Note that LFR-D does not allow to combine different event types in the same network. So, each generated network includes one event type. Each event, except hide-appear, occurs at each time slice. For hide-appear, some communities are hidden at a randomly selected time slice and the same communities reappear at some later time slice.
 \begin{table*}
  \caption{Values of the attribute-related parameters for each experiment} \label{table:experiments}
  \begin{center}
    \tabcolsep = \tabcolsep
    \begin{tabular}{p{0.19\textwidth}|p{0.19\textwidth}|p{0.19\textwidth}|p{0.32\textwidth}}
    \hline\hline
     \textbf{Model Parameters} & \textbf{Experiment 1}& \textbf{Experiment 2}& \textbf{Experiment 3} \\ 
     \hline\hline
 	Number of attributes  &  $\{\mathbf{1,3,5}\}$  &  $\{3\}$  &  $\{3\}$ \\ 
 	Values of attributes &	$[0,9]$  &  $[0,9]$  &  $[0,9]$ \\ 
 	Distribution Type  &  $\{$degenerate$\}$  &  $\{$degenerate$\}$  & $\{$\textbf{degenerate, binomial, uniform, power $\}$	}\\ 
 	Evolution percentage  &  $\{0\}$  &  $\{\mathbf{5,20,50,100}\}$  &  $\{5\}$ \\ \hline
 \textbf{Aim} &  Descriptor number  &  Descriptor stability &  Descriptor homogeneity \\ 
 \hline
    \end{tabular}
  \end{center}
 
 \end{table*}
\subsubsection{Generating Nodal Attributes}
Neither the original LFR model nor its extension LFR-D are able to generate networks with nodal attributes. Moreover, we could not find any study focusing on the generation of attributed networks in the literature. This might be due to the fact that the number of attributes, their domains and distribution over the network and the communities could be very system-specific. In order to fill this absence, we propose a relatively simple yet flexible model extending that of Greene \textit{et al}., able to associate attributes to nodes. Our goal here is not to deliver a realistic model, but rather to produce some controlled data which we will use to evaluate the performances of our framework. 

Our LFR-DA model allows to control the set $A$ of generated attributes through $4$ different parameters. First, the user must specify $|A|$, the number of attributes to be generated for all nodes. Second, it is necessary to define the domain of each attribute, i.e. the different values $\mathcal{D}_a$ an attribute $a$ can take. It is specified through two integers representing its upper and lower bounds, noted $min_a$ and $max_a$, respectively. For instance, if $min_a=2$ and $max_a=5$, the attribute $a$ can take the values $\mathcal{D}_a=\{2,3,4,5\}$. Third, the evolution percentage $q$ represents the proportion of nodes whose attribute values will change at each time slice.

Fourth, one must select a desired distribution type for each attribute, which describe how its values are distributed over a given community. This distribution parameter noted $h$ is categorical, and can be either \textit{Degenerate} (all nodes take the same randomly picked value), \textit{Binomial} (a significant proportion of the nodes take a randomly chosen value, the remaining ones take slightly different values), \textit{Power} (the values follow a power-law distribution, which means most of them have a very small value and only a few take a very large value), and \textit{Uniform} (all the domain values are evenly represented).
	
The procedure to generate the attributes is very simple and flexible. For each attribute $a$, we first generate its domain by respecting the specified bounds $min_a$ and $max_a$. Then, for each community of the first time slice, we generate the specified number $|A|$ of attribute values for each node, by respecting the specified distribution $h$. For the following time slices, for each community, we randomly select some nodes by respecting the percentage $q$ and change all their attribute values randomly, by respecting the domains and distribution of each attribute.

\subsection{Experimental Results}
In this section, we focus on the effect of descriptors on our method, especially in terms of scalability. Without loss of generality, we focus only on the attributes, but our conclusions are still valid for the topological measures. We perform three different experiments to analyze the behavior of mining CFS and calculating their emergence, on networks including nodal attributes. We use LFR-DA to generate network containing $5000$ nodes over $10$ time slices. For simplicity matters, all the attributes follow the same distribution. The values of the attribute-related parameters are described in Table \ref{table:experiments}. For all three experiments, we used the same range for the attributes: $min_a=0$ and $max_a=9$ (i.e. 10 possible discrete values).

In the first experiment, we studied the effect of the number of attributes $|A|$. The attribute values are generated so that they are completely homogeneous inside each community (\textit{degenerate} distribution). Moreover, these values are not affected by time. In the second experiment, our aim was to see how our framework is affected by changes in the evolution percentage of attributes $q$. For this reason, we fixed $|A|=3$, again with a \textit{degenerate} distribution, and modified only $q$. Finally, in the third experiment, we changed the distribution type $h$, whereas the number of attributes and the attribute evolution percentage were fixed at $3$ and $5$, respectively.

For each experiment, we extracted the node sequences, added the LFR-generated community to get the enlarged node sequences, and built a sequence database. We then applied CloSpan with $min_{sup}=10$ nodes, and processed the growth rates of the identified CFS.
\begin{figure*}
	\center
   \includegraphics[width=0.6\textwidth]{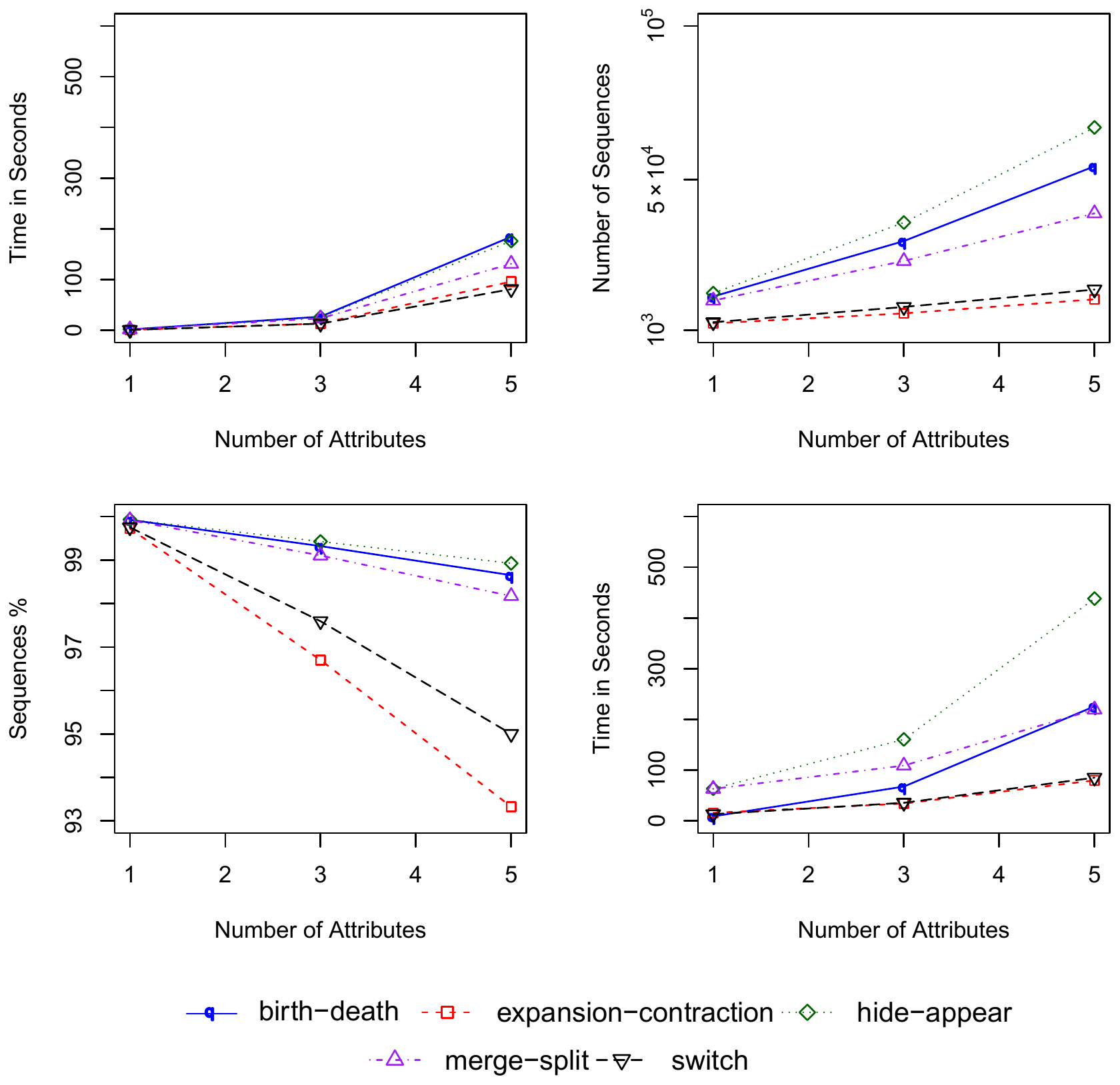}
 \caption[Results of Experiment 1.]{Results of Experiment 1: execution time of CloSpan (top-left), total number of sequences found by CloSpan (top-right), percentage of community-related sequences (bottom-left) and execution time for emergence computation (bottom-right), as functions of the number of attributes $|A|$. Colors represent the different types of community-related events.}
 \label{fig:exp1}
 \end{figure*}
\subsubsection{Effect of the Number of Attributes}
The effect of the number of attributes is illustrated by the four plots from in Fig. \ref{fig:exp1}. Here, we see that increasing the numbers of descriptors makes both mining CFS (top-left plot) and computing the growth rate of community-related sequences (bottom-right plot) more difficult. The number of descriptors obviously affects the number of candidate sequences, thus, it naturally impacts the execution time of CloSpan. The number of CFS increases with the number of descriptors (top-right), as expected. We remark that most of those sequences (at least $93\%$) are community-related (bottom-left). This is consistent with the fact all the nodes of the same community have the same attribute values in this experiment (degenerate distribution). So, a community-independent sequence is rarely closed, because there often exists a community-related super-sequence with the same support. The percentage of community-related sequences decreases when the number of attributes increases, though. This means the number of community-related sequences increases slower than that of all CFS. The execution time of emergence computation clearly increases with the number of attributes, like for CloSpan, and even faster. Indeed, it directly depends on the number of community-related sequences. This confirms our analytical estimation of the algorithmic complexity of this process, which depends on the number of nodes $n$, the number of time slices $\theta$ and, most of all, the number of sequences fetched by CloSpan $r$.

If we order the execution times in function of the type of evolution used for network generation (represented by colors in Fig. \ref{fig:exp1}), we get (in descending order): hide-appear, birth-death, merge-split, expansion-contraction and switch. So, when the community evolution undergoes small-scale events (expansion-contraction and switch), mining the sequences and computing emergence requires less time. These events do not cause the appearance or disappearance of communities, so they have no effect on the number of communities present in a time slice, unlike large-scale events. Communities are considered as an additional item in our sequence database, so less communities means a smaller number of possible sequences, and consequently shorter processing times. 

\begin{figure*}
	\center
  \includegraphics[width=0.6\textwidth]{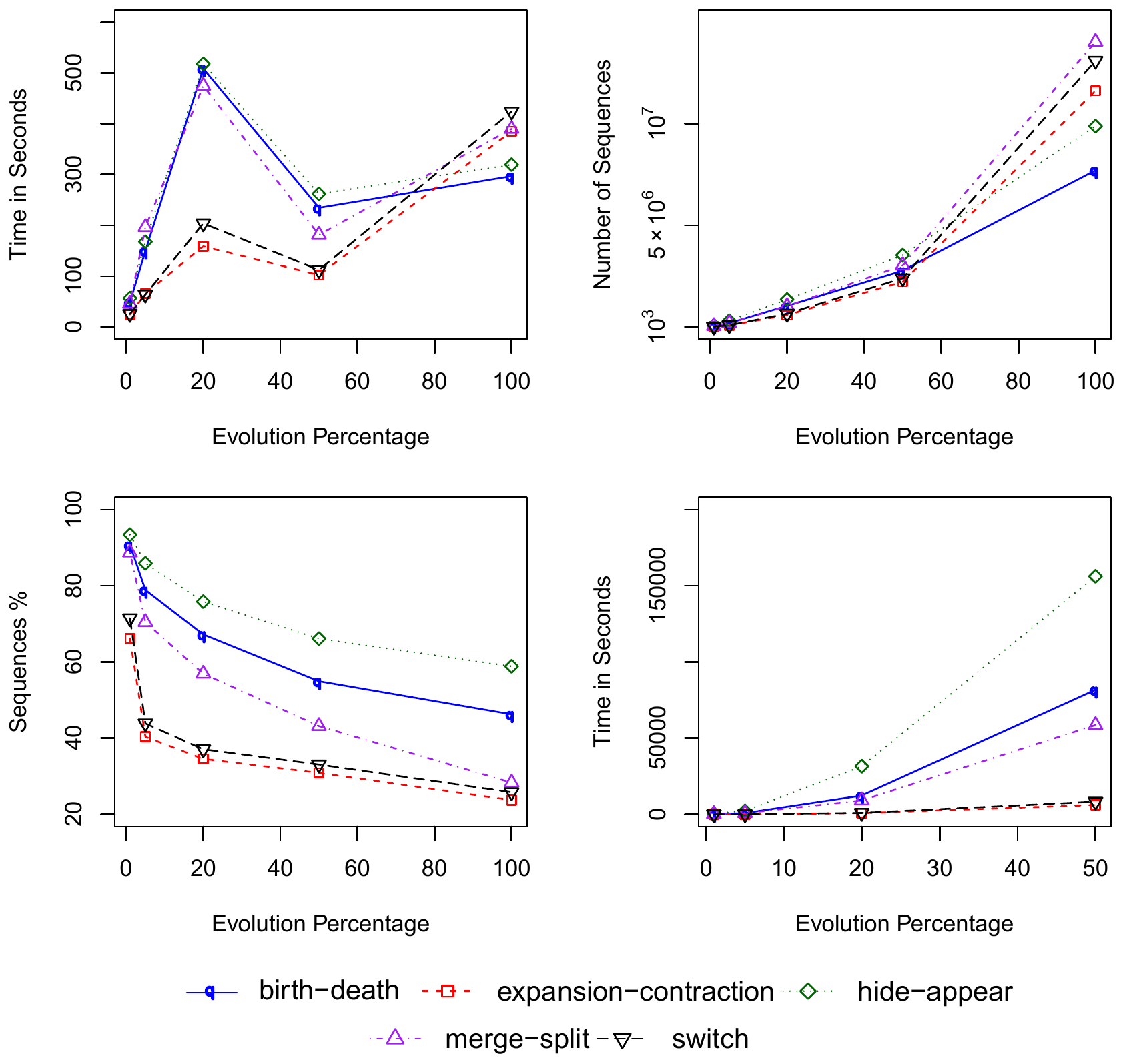}
\caption[Results of Experiment 2.]{Results of Experiment 2: execution time of CloSpan (top-left), total number of sequences found by CloSpan (top-right), percentage of community-related sequences (bottom-left) and execution time for emergence computation (bottom-right), as functions of the evolution percentage $q$. Colors represent the different types of community-related events.}
\label{fig:exp2}       
\end{figure*}
\subsubsection{Effect of the Attribute Stability}
The effect of the attribute stability is shown in Fig. \ref{fig:exp2}, with a plot layout similar to that of Figure \ref{fig:exp1}. The fastest CloSpan processing (top-left plot) is obtained for $q=5$, i.e. when the attribute values of $5\%$ of the nodes change at each time slice. The computation time increases very irregularly for higher values of $q$. When $q=20$, the execution time is the highest for large-scale events, whereas for small-scale events, it is when $q=100$. Note that modifying $q$ not only causes a change in the attribute evolution, but as a side effect, it also affects the distribution of attribute values inside the communities. The higher $q$, the earlier the communities become heterogeneous, in terms of attribute distribution. When $q\geq 50$, more than half the nodes see their attributes randomly changed at each time slice. The distribution therefore becomes more and more uniform (and the attributes values heterogeneous). 

A higher heterogeneity in the distribution should affect CloSpan negatively in terms of computational time, because it leads to more candidate CFS to generate. However, this effect is limited by the small number of values in our attribute domains ($|\mathcal{D}_a|=10$), hence the execution times observed at $q=50$. The increase at $q=100$ is due to an increase in the number of community-independent candidate patterns over the whole network, as shown by the plots of total number of patterns (top-right) and percentage of community-related patterns (bottom-left). 

\begin{figure*}
	\center
  \includegraphics[width=0.6\textwidth]{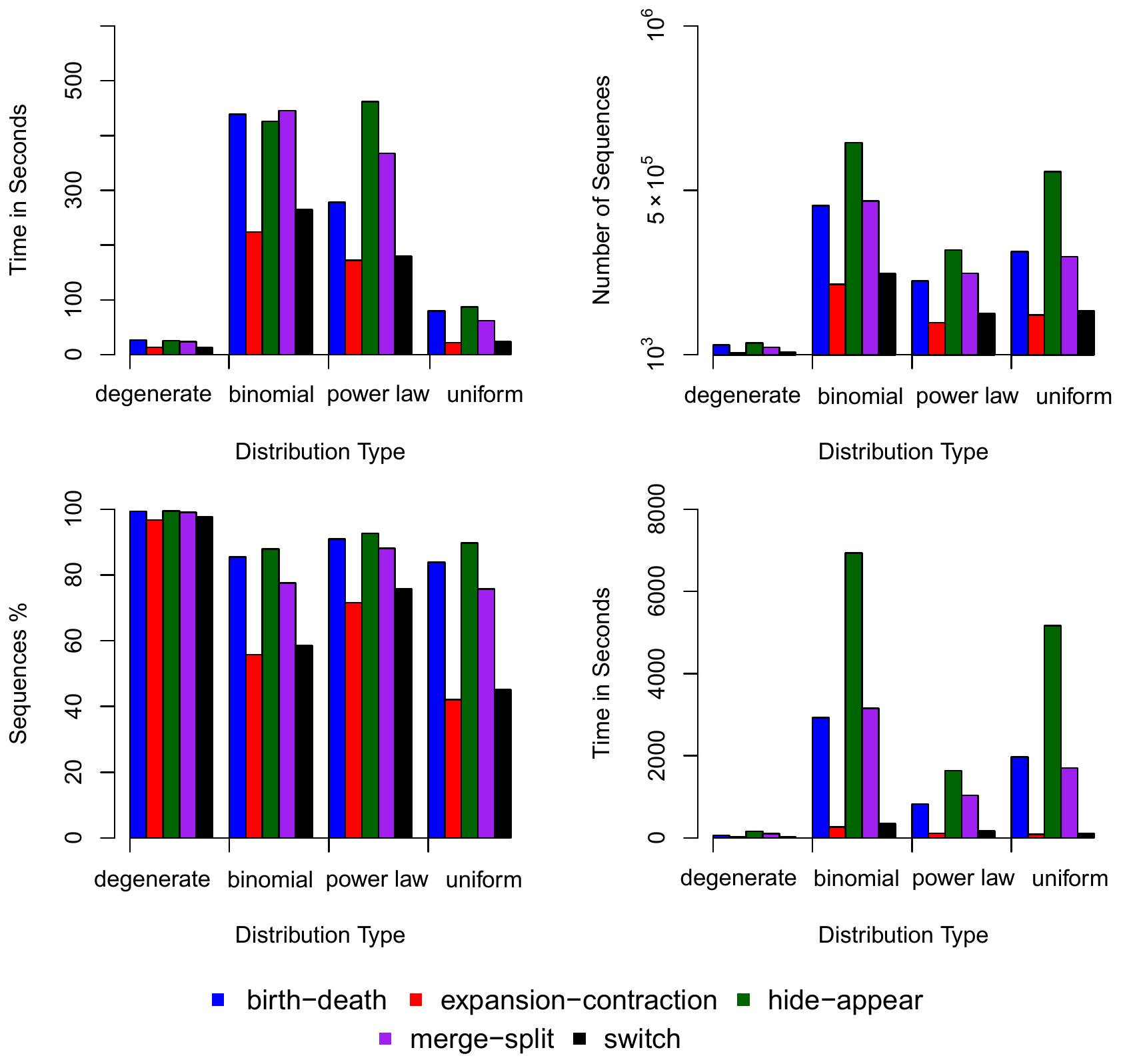}
\caption[Results of Experiment 3.]{Results of Experiment 3: execution time of CloSpan (top-left), total number of sequences found by CloSpan (top-right), percentage of community-related sequences (bottom-left) and execution time for emergence computation(bottom-right), as functions of the attributes distribution $h$. Colors represent the different types of community-related events.}
\label{fig:exp3}
\end{figure*}
Like for the attribute number $|A|$, the execution time of emergence computation increases with the evolution percentage $q$. Indeed, it directly depends on the number of community-related sequences.The most difficult event type is hide-appear, which is consistent with the fact it also leads to more distinct community-related sequences. It is followed by birth-death, merge-split, expansion-contraction and switch. The time necessary to handle the community-related sequences for small-scale events is much smaller than for large-scale events. It seems also stable, thus not affected by the increase in the percentage of modified nodes. This is due to the fact there are fewer community-related sequences for these two events (bottom-left).

\subsubsection{Effect of the Attribute Distribution}
Fig. \ref{fig:exp3} presents our results relatively to the effect of attribute distribution. The plots are organized similarly to the previous figures. CloSpan is the fastest (top-left plot) when the attribute values are completely homogeneous (degenerate distribution) or completely heterogeneous (uniform distribution). Because of the finite size of the attribute domains, a uniform distribution leads to communities containing groups of similar nodes of approximately the same size. If the community is large enough, the trends of these node groups will be represented as different super-sequences of the same community sequence. If it is too small, the $min_{sup}$ threshold will not be reached, and our tool will not detect the sequences. However, since we use the same attribute values for all communities in this experiment, the total size of these small node groups over the whole network is large enough to exceed the threshold.

When the attributes values are relatively homogeneous (binomial distribution), the execution time is at its highest. In this case, there are many nodes having the same value for a given attribute, and a few ones with different values. It seems these few numbers of nodes with different values are causing the identification of many extra candidate sequences (top-right plot). For both binomial and power-law distributions, we see a distinction between the processing of small-scale events, which is faster than for large-scale ones.

The results obtained in these three experiments validate and complete the analysis presented in Section \ref{sec:method} regarding the computational complexity of our tool. They highlight the fact the computing time is also affected in practice by the number of attributes (which directly influences $r$, the number of CFS identified by CloSpan) and their distribution. Moreover, the process becomes harder when the studied network undergoes large changes. This holds for both the community structure and the attribute values.

\section{Validation on Real-World Networks}
\label{sec:Valid}
We apply our method on two dynamic attributed networks modeling real-world systems: the first is a coauthorship network from the DBLP website, and the second is a friendship network from the LastFM service. 

We treated both of them in exactly the same way. First, we removed the isolated nodes, which do not have any interest in this context. Then, as explained in Section \ref{subsubsec:CommunityDetection}, we used Incremental Louvain to detect communities. The obtained modularity is larger than $0.6$, for each time slice and on both networks, which is a sign of well-separated communities. The preprocessing of the descriptors corresponds to what we described in Section \ref{subsub:dataprep}, including the clustering of the $9$ topological measures into $3$ attributes representing the position of the node at the local (degree and transitivity), intermediate (community-related measures) and global (centralities and eccentricity) levels. The preparation of the attributes is system-dependent, and is therefore described later. 

We built the sequence database $M$ from the resulting descriptors and communities, and applied the process described in Section \ref{subsec:minedb}. We interpreted all the identified characteristic patterns, but it is obviously not possible to present them all in this article. We therefore focus on the most representative, interesting and/or relevant ones, in terms of interpretation of the communities, in order to give an example of how our method can be used and how to comment its outputs.

\begin{table*}
\caption{Characteristic patterns detected for two DBLP communities.}
\label{tab:DBLPPatterns}       
\scriptsize
\center
\begin{tabular}{l|l|l|l|l|l}
\hline
Time  & Community  & Pattern & Support & Growth & ID  \\
Slice &  Size &  &  &  Rate  &   \\

\hline
6 & 120 & \textit{\{high embeddedness\}\{SDM=1\}\{high betweenness, low closeness, high participation coeff.\}}
& 40 & 6.86  & 1\\

 & &\textit{\{high betweenness, low closeness, high participation coeff.\}} & & &\\
 \cline{3-6}

 &  & \textit{\{high betweenness\}\{high embeddedness, ICDM=1\}}& 40 & 6.00 & 2 \\
  &  & \textit{\{total conference between 1 and 5\}}&  & & \\ 
 \cline{3-6}
 
 &  & \textit{\{high embeddedness\}\{TKDE=1, total journal between 1 and 5\}}& 40 & 4.00 & 3\\ 
  &  & \textit{\{high embeddedness, total journal between 1 and 5\}}&  & &\\ 
   &  & \textit{\{total journal between 1 and 5\}\{total journal between 1 and 5\}}&  & &\\ 
\cline{3-6}
 \hline 
8 &  113 & \textit{\{ILP=1\} \{high betweenness, low closeness, high embeddedness\}} & 40 &  59.08 & 4 \\
  &&&& &\\  \hline

\end{tabular}
\end{table*}

\subsection{DBLP Dataset}
\subsubsection{Data and Preprocessing}
DBLP is a bibliographic Website focusing on Computer Science works. We selected the dynamic co-authorship network of Desmier \textit{et al}. \cite{Desmier2012}, extracted from the DBLP database. Each one of the $2145$ nodes represents an author. Two nodes are connected if the corresponding authors published an article together. Each time slice corresponds to a period of five years. There are 10 time slices in total, ranging from 1990 to 2012. The consecutive periods have a three year overlap for the sake of stability. For each author, at each time slice, the database provides the number of publications in $43$ conferences and journals. We used this information to define $43$ nodal attributes corresponding directly to the individual conferences and journals, and $2$ additional ones representing the total number of conference and journal publications, respectively. We consequently have a total of $45$ nodal attributes. 

The conference/journals we selected are related to the subjects of database, data mining, knowledge discovery, information retrieval or artificial intelligence. To lighten the computational load, we decided to bin their values. For a journal/conference publication, we determined $5$ categories, corresponding to the number of publications $1$, $2$, $3$, $4$ and greater or equal to $5$. For both attributes representing the total conference and journal publications, we defined $5$ categories as well, but they correspond to different ranges: $[1;5]$, $]5;10]$, $]10;20]$, $]20;50]$ and $]50;\infty [$. These thresholds were determined according to our knowledge of the domain. For the mining step, we used $min_{sup}=21$ nodes, $min_{gr}=1.00$ and $max_{seq}=5 $. We obtained $1106108$ closed frequent sequences with these parameter values, $19922$ of which ($\simeq 1\%$) were community-related. The total execution time for finding the CFS was approximately $400$ seconds, whereas it was close to $6000$ seconds for the post-processing. Although the DBLP network has much more descriptors than our artificially generated networks, we see that these execution times are consistent with the experimental results obtained on the LFR networks for the same numbers of result patterns.

\subsubsection{Interpretation}
The communities of the DBLP network are very dynamic and change much through time. In the beginning of the considered time period, the communities are in general smaller. As time goes by, small communities tend to merge into larger ones, appearing around $t=7-9$. Descriptor-wise communities are not described by a single characteristic pattern: several ones are required to get a sufficient coverage. Communities are not homogeneous in terms of conference or journal, their members publish on several different scientific platform. 

\begin{figure*}
	\center
  \includegraphics[width=0.75\textwidth]{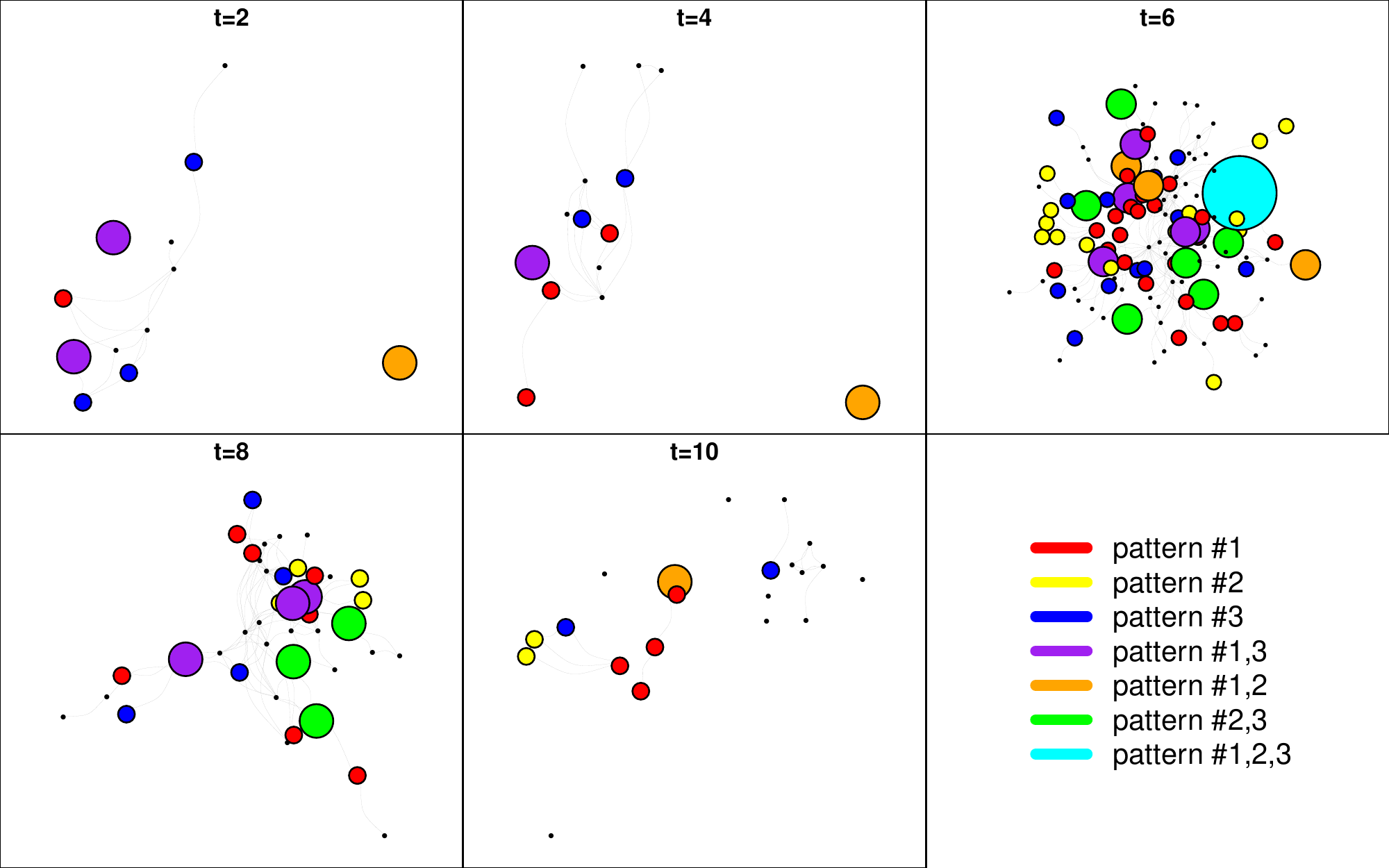}
\caption[DBLP Community]{Evolution of a community of $120$ nodes appearing at $t=6$ in the DBLP data. Nodes not following any characteristic patterns are represented in black. Red, yellow and blue nodes follow one of the three characteristic patterns. Orange, purple and green nodes follow 2 patterns and the cyan node follow all three characteristic patterns. The node size represents the number of supported characteristic patterns. Pattern-related details are given in Table~\ref{tab:DBLPPatterns}.}
\label{fig:DBLPComm63110}
\end{figure*}

As an example, we focus on a community appearing at $t=6$ (i.e. 2000-2004) and containing $120$ nodes. Note that, except at $t=6$, these $120$ nodes belong to several different communities. We have found $3$ characteristic patterns for this community, which are listed, with their interpretation, support and growth rate, in Table~\ref{tab:DBLPPatterns}. As mentioned before, the topological descriptors have been discretized by means of a cluster analysis. Instead of describing the patterns in terms of meaningless cluster numbers, we name them using their most characteristic feature (e.g. \textit{high embeddedness} for the group of community-related measures). The evolution of the $120$ nodes at $t=2,4,6,8$ and $10$ is represented in Fig.~\ref{fig:DBLPComm63110}. Their colors represent which ones of the community characteristic patterns they support, and their sizes are proportional to the number of supported patterns. For readability matters, at each time slice, we show only the nodes belonging to the largest community.

As shown in the figure, most of the nodes do not belong to the same community in the first time slices. They gather together at $t=6$ time and split again later. This community is not homogeneous around one conference or journal, since we identified $3$ characteristic patterns involving different scientific platforms (SDM, ICDM and TKDE). However, all of them are related to data mining, so the community is thematically homogeneous. The topological information present in all three patterns describes nodes strongly belonging to their community and critical for information flow (high betweenness). Pattern \#1 shows that, after publishing in SDM once, the nodes get connections with many other communities (high participation coefficient). Considering the growth rate of the patterns \#1, 2 and 3, we conclude that these patterns are not highly emergent for this community. Investigating its reason, we see that in the network, there are other nodes with similar trends. Nevertheless, these other nodes are not neighboring the community of interest. 

Table~\ref{tab:DBLPPatterns} also shows the characteristic pattern for another community, appearing at $t=8$ and containing $113$ nodes. This pattern tells us the community contains a group of nodes which publish at ILP once, and then get high betweenness and embeddedness whereas their closeness gets low. As it is seen in the table, the growth rate of the pattern is very high, which means the nodes supporting it constitute a distinctive trend in the community.

\begin{table*}
\caption{Characteristic patterns detected for three LastFM communities.}
\label{tab:LastFMPatterns}       
\scriptsize
\center
\begin{tabular}{l|l|l|l|l|l}
\hline
Time  & Community  & Pattern & Support & Growth & ID  \\
Slice &  Size &  &  &  Rate  &   \\

\hline
1 & 105 & \textit{\{Miles Davis between 1-5\}} & 20 & 20.36   & 1\\

 & &\textit{\{high degree, high local transitivity, community non-hub, low participation coeff.\}} & & &\\
  & &\textit{\{high degree, high local transitivity, community non-hub, low participation coeff. } & & &\\
   & &\textit{  high betweenness, Miles Davis between 1-5 \}} & & &\\
\cline{3-6}

 &  & \textit{\{high degree, high local transitivity, Ella Fitzgerald between 1-5,  Chet Baker between 1-5\}}& 20 & 8.84 & 2 \\
 &  & \textit{\{high degree, high local transitivity\}}&  & &\\ 

\cline{3-6}
 \hline 
1 &  81 & \textit{\{community non-hub, low participation coeff.\}} & 15 &  14.97 & 3 \\
 &  & \textit{\{community non-hub, low participation coeff.\}}&  & &\\ 
  &  & \textit{\{Pink Floyd between 5-10\}}&  & &\\ 
    &  & \textit{\{community non-hub, low participation coeff. and The Beatles between 5-10\}}&  & &\\ 
     &  & \textit{\{community non-hub, low participation coeff.\}}&  & &\\ 
\cline{3-6}
 &  & \textit{\{The Beatles between 5-10\}\{The Beatles between 5-10\}\{Frank Sinatra between 1-5\}}& 15 & 7.23& 4\\ 
\hline
1 and 4 &  22 & \textit{\{community non-hub, low participation coeff.\} \{Pink Floyd between 5-10\}} & 15 &  6.21 & 5\\
\cline{3-6}
\hline
\cline{3-6}
\hline
\end{tabular}
\end{table*}

\subsection{LastFM Dataset}
\label{subsec:LastFM}
\subsubsection{Data and Preprocessing}
LastFM is a music Website that allows its members to register and listen to music online. It is also a social network platform, because the members can declare friendship relationships. In LastFM, members can join predefined groups related to their music tastes, and declare their participation to music-related events such as concerts. We extracted a network by focusing on the members of the \textit{Jazz} group, which is supposed to include users appreciating this type of music.

We took advantage of the LastFM API and of the Jazz group statistics provided by LastFM to retrieve its members and identify the highest rated artists and event types for this group. It is interesting to notice that, although all the members we are interested in are from the Jazz group, the selection of the most popular artists include acts from music types which are very different from Jazz, such as the Beatles or Pink Floyd. We nevertheless kept these artists, because their playing rate is very high for the group. Event types also span a wide range of music types, from Jazz to Folk. We tracked the activity of the Jazz group members for the year $2013$. We created a time evolving network containing $3478$ nodes (each one representing a users from the Jazz group) and including $12$ time slices (each one representing a $1$ month period).

We defined $37$ nodal attributes based on the users' music tastes: $21$ of them represent the most popular artists in the Jazz group, and the $16$ others correspond to the types of events the members can join. For each time slice, the attribute values correspond to the total numbers of times a given user listened to a given artist, and the number of times he joined an event of a given type, respectively. We put a link between two nodes if two conditions were simultaneously true: 1) both considered users listened to at least one common artist for a specific period of time, and 2) they declared a friendship relation on the LastFM platform. 

Once the network was created, we processed the required topological measures and clustered them as we did for the DBLP network. The attribute values were binned. For artist attributes, we determined $5$ intervals regarding the numbers of listenings: $[1;5]$, $]5;30]$, $]30;90]$, $]90;200]$ and $]200;\infty[$. Regarding the event types, we defined $4$ intervals: $[1;4]$, $]4;7]$, $]7;11]$ and $]11;\infty[$. These thresholds were determined according to our knowledge of the domain. For mining, we used $min_{sup}=15$ nodes, $min_{gr}=1.00$ and $max_{seq} =5$. With these parameter values, we found $632037$ closed frequent patterns, $44631$ of which ($\simeq 7\%$) were community-related. The total execution time for finding the CFS was close to $200$ seconds, whereas ut was approximately $7000$ seconds for post-processing. Like for DBLP, we see that the LastFM execution times are consistent with the results obtained on LFR networks.

\subsubsection{Interpretation}
The evolution of the LastFM communities is generally smoother than for DBLP. We observe relatively large communities at $t=1$ and $2$, and communities tend to evolve most of all through small-scale events (by opposition to large-scale events such as split or merge). We want to remind the reader that this LastFM network describes an evolution over one year, while the DBLP network represents a 18 years period. So not only the systems, but also the temporal scales are different. Table~\ref{tab:LastFMPatterns} shows the characteristic patterns obtained for three communities, which we use in the rest of this section to answer three questions regarding these data. 

Like for DBLP, we identified many community sequences and their characteristic patterns. We focus our comments on a few of them, in order to illustrate how our tool can be useful, through three questions the end user interested with these data is likely to ask himself.

\paragraph{Which topological descriptors matter?}
Table~\ref{tab:LastFMPatterns} first presents two characteristic patterns of a community present at the first time slice and containing $105$ nodes. The most emerging one is pattern \#1, and it represents a node group which listens to Miles Davis between one and five times in a time slice. Then, they take a high degree and local transitivity. Meanwhile they are non-hub in their community and have some external connections. However, these external connections are not distributed over many different communities, but rather concentrated on a few ones. In the future, these nodes keep this topological position, while also having a low betweenness and listening to Miles Davis between one and five times. Note there are other communities also interested in Miles Davis. By \textit{topological position}, we mean the set of topological features a node possesses, as described by the pattern it supports. 

Pattern \#2 refers to a node group which have a high degree and high local transitivity, while listening to Ella Fitzgerald and Chet Baker between one and five times. Those nodes keep the same topological position in latter time slices. Here, listening to Ella Fitzgerald and Chet Baker at the same time might be explained by the fact these two artists are vocal. Chet Baker is famous for his virtuosity on the Trumpet, but he also sings. Indeed, when we looked in detail at the LastFM data, we saw that the users from this group mainly listened to sung pieces such as \textit{My Funny Valentine}. 

For this community, we see common topological features regarding the community-related descriptors. The nodes are locally well-connected, but not as much at the level of the community structure. If we ignore these descriptors, the growth-rate of pattern \#1 drops to $3.87$ (instead of $20.36$), which shows the importance of these topological properties to distinguish the community from the rest of the network. 

\paragraph{How do non-Jazz listeners behave?}

The second community contains $81$ nodes and it is also present at $t=1$. Its characteristic patterns are listed in Table~\ref{tab:LastFMPatterns}. The topological position represented by Pattern \#3 describes a non-hub node with low participation coefficient, which persists through time. Regarding the attributes, these users listen to Pink Floyd $5$-$10$ times, then to the Beatles $5$-$10$ times. Although the artist they listen to changes, their structural features do not change at all, and moreover both artists are not Jazz acts. One interesting supplementary sequence shown as pattern \#4 refers to an emerging node group, which listen to the Beatles $5$-$10$ times for two time slices, and then to Frank Sinatra $1$-$5$ times too. The nodes supporting these two patterns have a 50\% overlap. More clearly, half of the nodes following pattern \#3 also follow pattern \#4. In fact, for each community, we had many overlapping patterns. We want to remind the reader that our pattern selection process picks up the most distant sequence from the already chosen ones among the closed and emergent sequences. If there is not a very distant sequence in terms of supporting nodes, it picks up the best choice, and this can result in some overlap.

Those two patterns include two clearly non-Jazz artists (The Beatles and Pink Floys), and one who could be considered as Jazz-relative (Sinatra). So, we can suppose the users from this community are not interested in Jazz, or only remotely. We did not find any other such community of non-Jazz listeners with our method. Pattern \#3 additionally gives some information regarding the topological features of the persons listening to these artists. It seems they do not have many connections outside of their communities, since they have a low participation coefficient for many time slices. 

\begin{figure*}
	\center
  \includegraphics[width=0.75\textwidth]{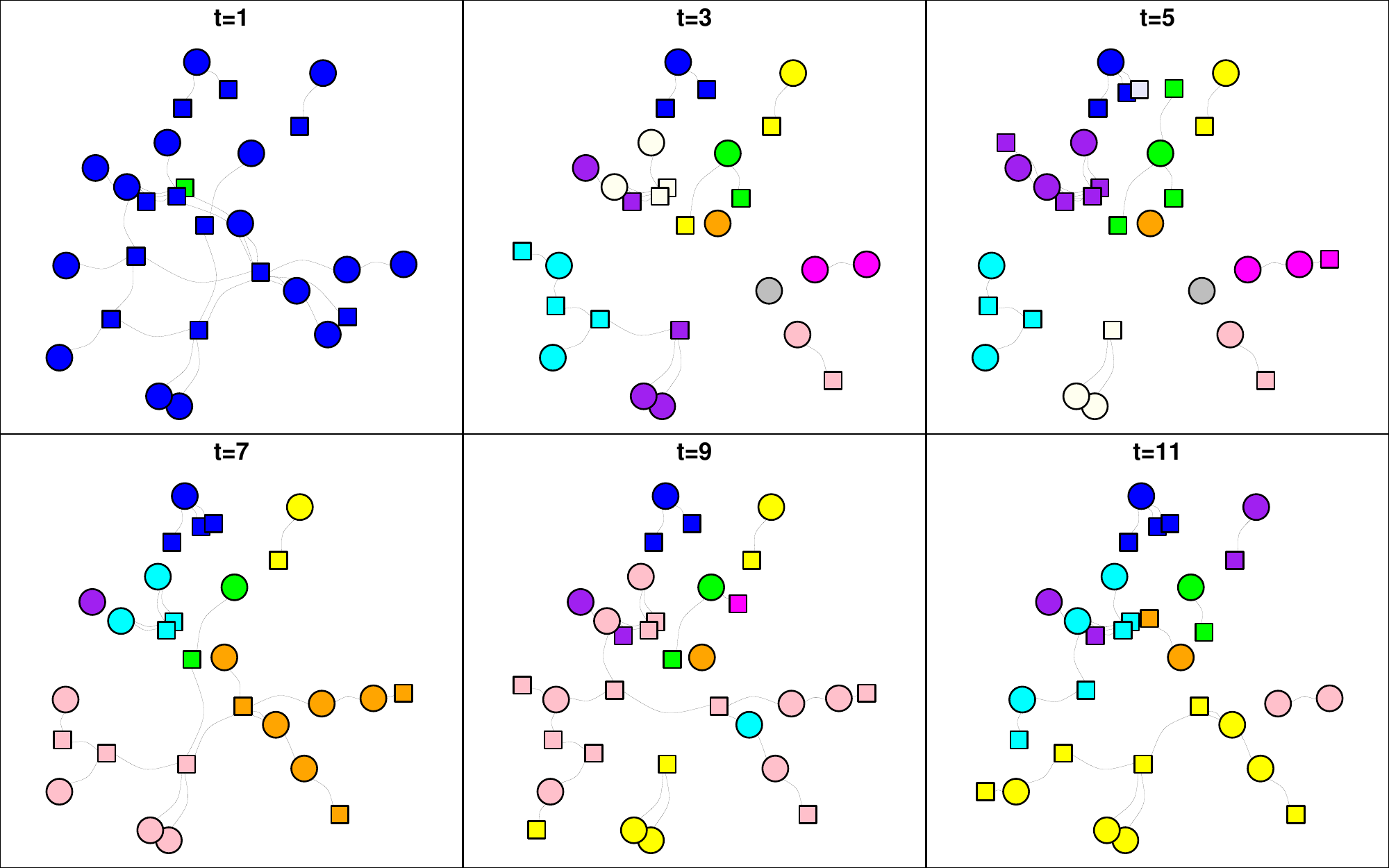}
\caption[LastFM Pattern]{Interconnection of the nodes following pattern \#3 from Table~\ref{tab:LastFMPatterns} (circles), and of their neighbors not following this pattern (squares), for $t=1$, $3$, $5$, $7$, $9$ and $11$. Each color corresponds to a specific community at a given time slice. The same color in two different time slices does not necessarily represent the same community.}
\label{fig:myPlotLastFMPattern}
\end{figure*}

We investigated the topology of this group for all the time slices, as illustrated in Fig.~\ref{fig:myPlotLastFMPattern}. Circles represent nodes supporting pattern \#3 and squares are neighbors not supporting it, while colors represent communities. Nodes supporting pattern \#3 may belong to different communities when considered at different time slices. But, independently from this observation, they do not have many connections outside of their communities. They are not central at all. Mostly, they belong to a tree-like structure, holding a non-hub role.

\begin{figure*}
	\center
  \includegraphics[width=0.75\textwidth]{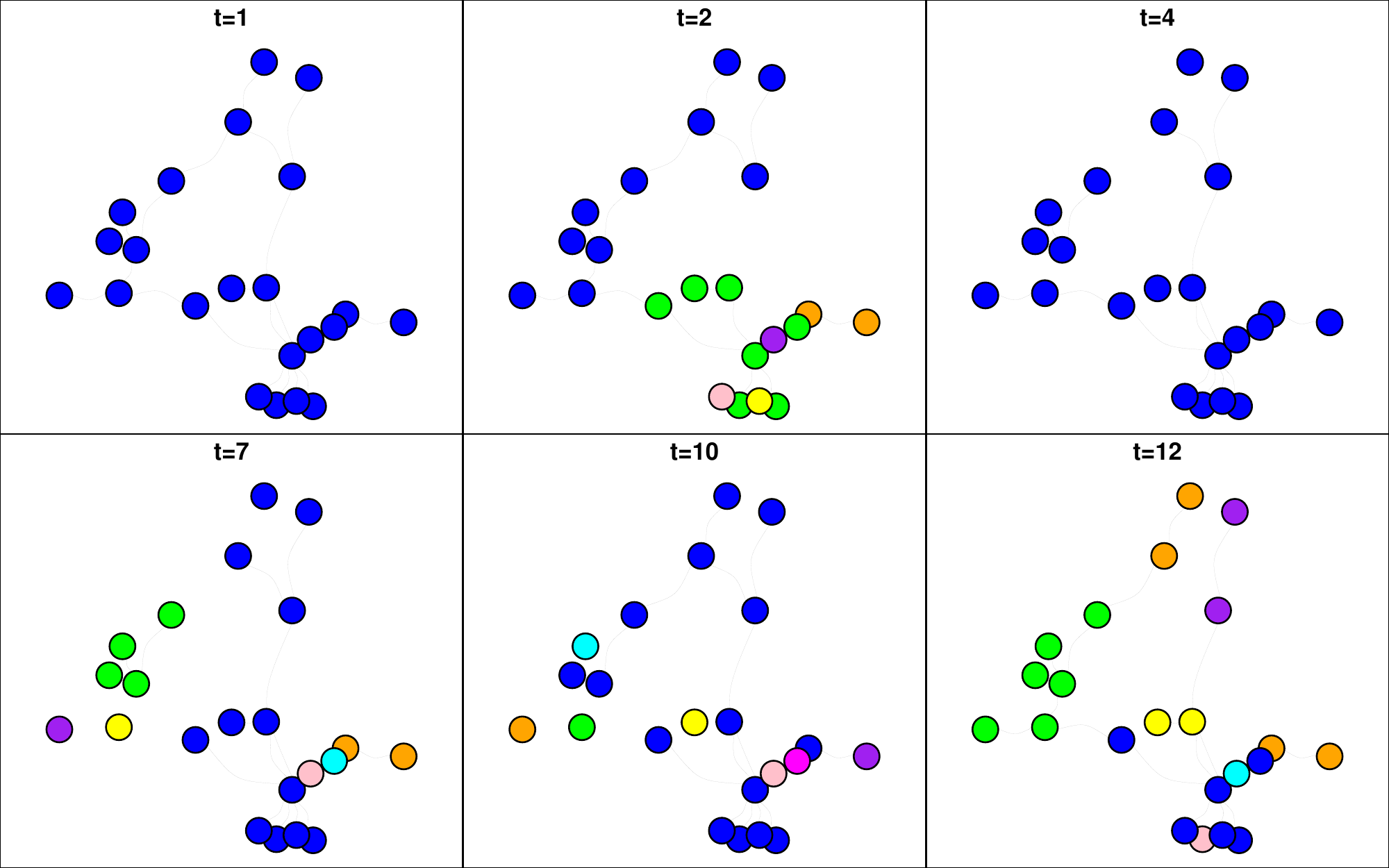}
\caption[LastFM Community]{Evolution of a group of $22$ nodes at $t=1$, $2$, $4$, $7$, $10$, $12$ in the LastFM data. Those nodes belong to the same community at $t=1$ and $4$. Each color corresponds to a specific community at a given time slice. The same color in two different time slices does not necessarily represent the same community.}
\label{fig:LastFMComm16020}       
\end{figure*}

\paragraph{What about community sequences?}
A very interesting fact concerning the second community is that some of its $81$ nodes from $t=1$ are still together at $t=4$. Thus, we also interpret the community sequence of $22$ nodes appearing at $t=1$ and $t=4$. The characteristic pattern for these $22$ nodes is shown as pattern \#5 in Table \ref{tab:LastFMPatterns}, and the evolution of their communities is illustrated by Fig.~\ref{fig:LastFMComm16020}. In the figure, we show the connections of those $22$ nodes for $t=1$, $2$, $4$, $7$, $10$, and $12$, which constitute this community sequence. The node colors correspond to different communities, and the same color in two different time slices does not necessarily match the same community, since communities undergo large-scale events, as explained in Section \ref{sec:overview}. 

Pattern \#5 describes a topological position consisting in being non-hub and having a low participation coefficient. Attribute-wise, these users listen to Pink Floyd $5$-$10$ times. This trend is very similar to that of the larger community we discussed just before (patterns \#3 and 4). This smaller sequence shows us that some nodes with similar interest keep their connections for several time slices, even if the rest of their community changes. The support of pattern \#5 shows this smaller community is more homogeneous than the larger community. The same observation can be done for many other communities in our results as well. This could be interpreted as the presence of a certain hierarchical structure in the network, defined in terms of both topology and attributes (i.e. nested interests). However, a more thorough study shows this type of hierarchy is not stable through time. For this reason, it is more accurate to consider communities have one or several cores, corresponding to nodes supporting long sequences, i.e. evolving together for a long time. These cores can overlap and switch communities. They are joined punctually by other cores or peripheral nodes to form short-termed communities. The way those node groups move in the community structure could be studied to understand how trends propagate through a network.

\subsection{Discussion}
Let us conclude the analysis of the real-world networks with a comparison of the obtained results. Regarding the time performance, we can say that although both networks have smaller sizes and much more node descriptors than the generated artificial networks considered in section \ref{sec:Valid}, the percentage of community-related sequences we identified is larger by less than $\%7$. This makes the execution time of the growth-rate computation lower than for most of these artificial networks. One reason for this observation can be the minimum support threshold we used, which puts a limit to the community sizes. As a result, there are fewer community-related sequences. About the execution time of the CFS mining, we know the numbers of node descriptors affect it. So we expected it would run slower for our real-world networks, which contain many descriptors. However, our results show that there is no significant difference between the performances obtained on artificial and real-world networks. One reason for this could be the heterogeneity of the descriptor values over the nodes: it leads to the generation of less candidate sequences, and CloSpan consequently works faster than expected.

The community structure of both networks changes very much at each time slice, especially for DBLP. The communities in this network are getting larger with time, whereas for LastFM, their sizes remain relatively constant. For a large part of the communities, the most supported sequences involve mainly topological descriptors. For DBLP, inside a given community, almost all the nodes ($\sim90\%$) have the same topological situation. For most of the communities, this situation is: being non-hub and having a low participation coefficient. But certain communities are also characterized by non-central, low degree nodes. For LastFM, we also find purely topological sequences, in particular: being non-hub and having a zero or low participation coefficient. But a number of sequences also reflect an interest towards specific Jazz artists. The proportion of nodes following the most supported sequences is not as high as for DBLP ($\sim70\%$). 

It is worth noting the combination of topological and attribute-based descriptors help interpreting the patterns: changes in the topological aspect can help explaining changes in the attributes, and vice-versa. For example, in Table~\ref{tab:DBLPPatterns}, pattern \#1 displays a changes in the topological features: the nodes have at first a high embeddedness, which means most of their neighbors are in the same community. Then, they publish in SDM, and then have a high participation coefficient, meaning they get connected to (many) other communities: this can correspond to the beginning of new collaborations started after having published in a prestigious conference. On the contrary, node groups which are more embedded in their communities keep publishing in the same conferences.

For both networks, we cannot claim the communities are constituted of nodes gathering around a single interest. The characteristic patterns show us several distinct trends exist simultaneously in the same community. So, the communities are not homogeneous in terms of patterns. However, note the term homogeneous applies here to the descriptors constituting the patterns: different descriptors, especially nodal attributes, can actually have very close meanings. For instance, we showed a community of DBLP was characterized by several distinct patterns all related to journals and conferences from the Data Mining filed, i.e. thematically very similar.

The existence of several trends for the same communities means we have distinct subgroups of nodes supporting them. These subgroups are mobile, in the sense they switch from one community to the other. This is related to our previous remark regarding the fact the community structures of these networks are significantly changing from one time slice to the other. This can be a feature of the studied system, but it can also be an artifact of the community detection method, so this point should be considered carefully when applying our method.

\section{Conclusion and Perspectives}
\label{sec:conclusion}
In this work, we tackled the task of interpreting communities detected in dynamic attributed complex networks. We first formalized it as a data mining problem, then proposed a method to solve it. It is based on a sequence-based representation of the information, allowing to store simultaneously the topological information, the node attributes, the community structure and the temporal dimension. Our method takes advantage of this representation to search for frequent closed emerging sequential patterns, in order to characterize each community. This step involves the use of the data mining tool CloSpan \cite{Yan2003}. 
To our knowledge, this is the first time the task of interpreting communities is formulated as a problem, moreover independently from the community detection task. Our goal was to overcome the limitations of the few existing studies \cite{Labatut2012,Tumminello2011,Lancichinetti2010} by proposing a systematic approach, taking into account the topological structure, the nodal attributes and the network dynamics. Our sequential representation has not been used for graphs before. The process we proposed to extract the most relevant patterns based on closed and emergent sequences \textit{relatively to the communities} is original. 

Using artificially generated networks, we studied how our method is affected by certain properties of the treated data. To this aim, we proposed an extension of an existing generative model \cite{Greene2010}, allowing to produce dynamic attributed networks possessing a community structure. These empirical results confirm the analytic expression we provided for the algorithmic complexity of our method, and the effect of the network size (both in terms of nodes and time slices) and of the number of detected closed frequent patterns (itself related to the number of considered descriptors). 
We also used two real-world networks to validate our method. The first is a DBLP network showing the evolution of co-authorship connections over 18 years. We identified the presence of several trends in most communities, corresponding to subgroups focusing on different journals and conferences. These scientific platforms are connected to the same field, though (e.g. Data Mining, Artificial Intelligence, etc.), which reveals a certain thematic homogeneity in a community. The second network is extracted from LastFM, and represents the evolution of Jazz listeners, their friendship relations and listening habits for the year 2013. 
For this network, we also characterized thematically certain communities, depending on their favorite artists. We also identified a group of users which actually do not listen to Jazz.

In the future, we plan to apply our method to the analysis of other types of networks, in order to explore further its characterization capabilities. A related point consists in identifying which descriptors are the most relevant to describe a given system. For those considered here, the community-based topological measures were very discriminant, but this is not necessarily true for other systems. The preprocessing of the descriptors is also a critical point we want to explore. The main limitation of our method is its computational costs, which is due to the pattern mining perspective we adopted. Identifying the most important descriptors or performing a more appropriate preprocessing would allow to reduce the search space, and therefore to eliminate some redundant CFS, to ease the identification of emerging patterns, and to make the postprocessing faster. We also plan to improve our method itself, noticeably by using hash maps to process emergence.


\bibliographystyle{ieeetr} 
\bibliography{references}
%
%

\end{document}